\documentclass[12pt,a4paper]{article}
\usepackage{amsmath} 
\usepackage{amssymb} 

\usepackage{cases}

\usepackage{graphicx}
\usepackage[left=2cm,right=2cm,top=2cm,bottom=2cm]{geometry}

\usepackage[bookmarks=true,colorlinks=true,citecolor=blue,linkcolor=red,urlcolor=magenta]{hyperref}
\usepackage{xcolor}
\usepackage[numbers,sort&compress]{natbib}

\usepackage{soul}
\sethlcolor{yellow}

\title{Neutron Moderation Theory Taking into Accout the Thermal Motion of Moderating Medium Nuclei}
\author{V.D.~Rusov\footnote{Corresponding author e-mail: siiis@te.net.ua}, V.A.~Tarasov, S.A.~Chernezhenko, A.A.~Kakaev, V.P.~Smolyar}
\date{}

\begin{document}
\maketitle

\begin{center}
    \textit{Department of Theoretical and Experimental Nuclear Physics, \\ Odessa  National Polytechnic University, Odessa, Ukraine}
\end{center}

\begin{abstract}
In this paper we present the analytical expression for the neutron scattering
law for an isotropic source of neutrons, obtained within the framework of the
gas model with the temperature of the moderating medium as a parameter. The 
obtained scattering law is based on the solution of the kinematic problem of
elastic scattering of neutrons on nuclei in the $L$-system in the general case.
I.e. both the neutron and the nucleus possess the arbitrary velocity vectors in
the $L$-system. For the new scattering law the flux densities and neutron
moderation spectra depending on the temperature are obtained for the reactor
fissile medium. The expressions for the moderating neutrons spectra allow
reinterpreting the physical nature of the underlying processes in the thermal
region.
\end{abstract}

\section{Introduction}
\label{sec01}

An important part of the theory of neutron cycle in nuclear reactors is the
theory of neutron moderation~\cite{Weinberg1961,Akhiezer2002,Galanin1971,
Feinberg1978,BartolomeyBat1989,Shirokov1998,Stacey2001}. A  neutron moderation
theory traditionally used in contemporary nuclear reactor physics was developed
in the framework of the gas model. This model neglects the interaction between
neutrons and the nuclei of moderating medium, although some unfinished attempts
were made to include the interaction between the nuclei of the moderating
medium, e.g. in \cite{Akhiezer2002,Galanin1971}. This traditional theory of
neutron moderation is based on the neutron scattering law which defines the
energy distribution  of the elastically scattered neutrons in the laboratory
coordinate system ($L$-system) (see the neutron scattering law e.g. in
\cite{Feinberg1978,BartolomeyBat1989,Shirokov1998}). The neutron scattering
law, in its turn, is based on the solution of a kinematic problem of neutrons
elastic scattering on the moderator nuclei\cite{Weinberg1961,Akhiezer2002,
Galanin1971,Feinberg1978,BartolomeyBat1989,Shirokov1998}. It should be noted
however that we use the term "scattering law" instead of the "elastic
scattering law", because this solution is generalized for all kinds of
scattering reactions (elastic and inelastic) during the formulation of the
moderating neutrons balance equation. For example, as it is known from the
neutron moderation theory, the moderating neutrons flux density is found as a
solution of the balance equation for the moderating neutrons, and depends on
the macroscopic scattering cross-section $\Sigma_s$ in case of the moderator
without neutron absorption, and on the total macroscopic cross-section
$\Sigma_t$ in case of the moderator with neutron absorption 
(see~\cite{Weinberg1961,Akhiezer2002,Galanin1971,Feinberg1978,
BartolomeyBat1989,Shirokov1998} and sections~\ref{sec04}-\ref{sec07} below).
Let us remind that $\Sigma_s = \Sigma_{el} + \Sigma_{in}$ where
$\Sigma_{el} = \Sigma_p + \Sigma_{rs}$ is the macroscopic elastic scattering
cross-section, $\Sigma_p$ is the macroscopic potential scattering
cross-section, $\Sigma_{rs}$ is the macroscopic resonance scattering
cross-section, $\Sigma_{in}$ is the macroscopic inelastic scattering
cross-section, and $\Sigma_t = \Sigma_s + \Sigma_a$, where $\Sigma_a$ is the
macroscopic absorption cross-section for the moderating medium
(e.g.~\cite{Shirokov1998}).

The kinematic problem of an elastic neutron scattering on a nucleus in the $L$-system is a two-particle kinematic problem and may be solved exactly. Still, a neat and compact analytical solution of such problem may be obtained only in the case of the nucleus resting in the $L$-system before scattering. In the general case, when both the neutron and the nucleus have arbitrary velocity vectors in the $L$-system before scattering, the solution of this problem is a set of cumbersome expressions. This is because of the fact that an intermediate solution including the cosines of the angles between the nucleus and neutron velocity vectors is located in the C-system, and is rather lengthy by itself. The final solution of the problem in the $L$-system requires a transformation of the cosines from the C-system to $L$-system. And this requires several more relations transforming the unit vectors from the C-system to the $L$-system. Therefore, reduction of the solution to a single analytical expression makes no sense  in this case. However, in this case it is possible to build a computational algorithm to obtain the solution via computer calculation. This approach is implemented in modern Monte Carlo codes, e.g.  MCNP4, GEANT4 and others, which let one calculate the moderating neutron spectrum even for the moderators with absorption. It should be noted that these Monte Carlo codes are used for the majority of today's practical calculations involving the neutron spectra.

Still, the traditional theory of neutron moderation is based on the above
mentioned analytical solution obtained for the case of the nucleus resting in
the $L$-system before scattering. I.e. the traditional theory neglects the heat
motion of the moderator nuclei, which is acceptable if the neutron energies are
much higher than the thermal motion energy of the nuclei. As a consequence, the
neutron scattering law and the following analytical expressions for the Fermi
spectrum of the moderating neutrons do not contain the temperature of the
moderating medium. In order to cover  this significant gap in the theory, the
only thing suggested until today was to complement the Fermi spectrum of the
moderating neutrons with the Maxwell-type spectrum in the thermal energy range
artificially (in the sense that this was not obtained strictly from the
scattering law). Moreover, in order to form the Maxwell spectrum, it is
necessary to recalculate the temperature of the moderating medium $T$ into the
temperature of the neutron gas $T_n$ using the formula
$T_n = T \left[ 1 + 1.8 \frac{\Sigma_a (kT)}{\xi \Sigma_s} \right]$ (where
$\Sigma_a (kT)$ is the macroscopic absorption cross-section for moderating
medium and the neutrons of the energy $kT$, $\xi \Sigma_s$ is the moderating
power of the moderator for the 1~eV neutrons). According
to~\cite{Weinberg1961}, this formula was obtained as a numerical approximation
of the experimental spectra of several different types of nuclear reactors
available at that time, and  is still widely used in the reactor physics,
e.g.~\cite{BartolomeyBat1989,Shirokov1998,Stacey2001,Vladimirov1986,
Verkhivker2008,RusovEnergies2011,RusovVANT2011,RusovWJNST2013,RusovJMP2013,
RusovKhariton2010}. Let us also note that the multiplier before the second term
in brackets is often chosen by the developers depending on the reactor type,
e.g.~\cite{BartolomeyBat1989,Verkhivker2008}.

Because of the fission accompanied by the large energy release, the emission of neutrons and other particles, the nuclide composition change, the heat transfer, the radiation-induced defects dynamics (leading to the geometry change up to the complete destruction), reactor fissile medium is in the state of thermodynamic non-equilibrium. The same is true for any fissile medium with active chain reactions in it. I.e. the reactor fissile medium with the fission processes is an open physical system in the non-equilibrium thermodynamic state. Such system may be described within the framework of the non-linear non-equilibrium thermodynamics of the open physical systems. Such systems may include the non-equilibrium stationary states, which meet the Prigogine criterion -- the minimum of the entropy production (e.g. \cite{Prigogine2001,Bakhareva1976,Kvasnikov2003}). The realization and type of such stationary mode are known to depend both on the internal parameters of the system (internal entropy) and on the boundary conditions (boundary entropy flow). For example, the realization of the stationary state in a non-equilibrium system (hereinafter referred to as stationary non-equilibrium state) requires the constant boundary conditions (see e.g. \cite{Bakhareva1976}).

In our model of the neutron moderation in the fissile medium the following
simplifications are taken. Two thermodynamic subsystems are singled out from
the fissile medium -- the moderating neutrons subsystem and the moderator
nuclei subsystem. The subsystems are open physical systems interacting with
each other. Thus, according to the stated above, both of these systems are in
the non-equilibrium state. However, in our case we assume the moderator nuclei
subsystem to be near its equilibrium state because of its inertia relative to
perturbations and negligible influence of the neutron subsystem. This allows us
to introduce a temperature of the moderator medium. The neutron subsystem
remains non-equilibrium in our model, and the temperature of this subsystem is
not introduced. Let us emphasize that in order to construct the neutron
spectrum, the traditional approach operates with the concept of temperature of
the neutron gas, which indicates the usage of an additional simplification in
it -- the one that we refused to take. This fact, together with the
aforementioned, led the authors to a conclusion that there was currently no
robust and consistent theory of neutron moderation, and it was crucial to
develop such a theory.

The absence of the moderation theory results in numerous difficulties related
to the study of the reactor emergency modes, the development of the new
generation nuclear reactors such as the traveling-wave 
reactors~\cite{RusovSTNI2015,RusovPNE2015}, pulsed reactors, boosters,
subcritical assemblies \cite{Kolesov2007,Lukin2006,Arapov2010}, and the
investigation of the natural nuclear reactors such as
georeactor~\cite{RusovJGR2007}.

Let us also note that the expressions for the moderating neutrons flux density
obtained in the present paper, are the solutions of the equation describing the
process of the neutron moderation in a stationary state, which may set within a
non-equilibrium neutron system under certain conditions.

Thus, in the present paper, based on the solution of the kinematic problem of
elastic neutron scattering on a nucleus in the $L$-system in general case (when
both the neutron and the nucleus have arbitrary velocity vectors in the
$L$-system before scattering) we derive the analytical expression for the
neutron scattering law including the moderating medium temperature as a
parameter, for the case of an isotropic neutron source. We also obtain the
spectra of the moderating neutrons for different moderating media, which also
depend on the moderating medium temperature, and are true for virtually the
entire fission spectrum (except the energies comparable to the energy of
interatomic or intermolecular interactions in moderating medium, which requires
going beyond the gas model). The resulting expression for the spectrum of
moderating neutrons allows us to reconsider the physical nature of the
processes that determine the neutron spectrum in the thermal region.

\section{Kinematics of the elastic neutron scattering on a moderating medium nucleus}
\label{sec02}

We consider the elastic scattering of a neutron on the nucleus of moderating
medium. The moderating medium is described within the framework of gas model,
i.e. the nuclei are assumed to not interact with each other, but possess
certain kinetic energy due to their thermal motion.

\begin{figure}
\begin{center}
\includegraphics[width=9cm]{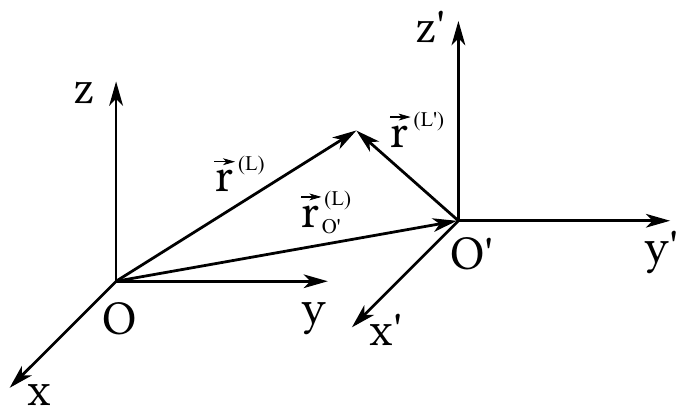}
\end{center}
\caption{Laboratory coordinate systems $L$ and $L'$.}
\label{fig01}
\end{figure}

At the very beginning the authors made an important assumption that the form of the desired solution of the kinematic problem of the neutron elastic scattering on a nucleus must be similar to  the one used within the traditional theory of neutron moderation. Therefore, in order for this solution to include the solution used by the traditional neutron moderation theory, as a particular case, it is convenient to introduce two laboratory systems (Fig.~\ref{fig01}):
\begin{itemize}
\item the resting laboratory coordinate system, referred to as the $L$-system;
\item the laboratory coordinate system moving relative to the $L$-system at a constant speed equal to the speed of the nucleus thermal motion in the moderating medium (this one is referred to as the $L'$-system).
\end{itemize}

So we consider the particular case when the spatial orientation of the coordinate axes of the $L$-system and the $L'$-system is the same, and the radius vector of the $L'$-system's origin in the $L$-system coincides with the radius vector of the moderating medium nucleus by which the neutron is scattered. I.e. the nucleus rests in the $L'$-system.

Let us introduce the following notation: $m_1 = m_n$ is the neutron mass;
$m_2 = m_N$ is the nucleus mass; $\vec{r}_1^{(L)}$ is the neutron radius
vector in the $L$-system; $\vec{r}_2^{(L)}$ is the nucleus radius vector in the
$L$-system; $\vec{r}_c^{(L)}$ is the radius vector of the center of mass in the
$L$-system; $\vec{r}_1^{(L')}$ is the neutron radius vector in the $L'$-system;
$\vec{r}_2^{(L')}$ is the nucleus radius vector in the $L'$-system;
$\vec{V}_{10}^{(L)}$ is the neutron speed in the $L$-system before the
collision with the nucleus; $\vec{V}_{1}^{(L)}$ is the neutron speed in the
$L$-system after the collision with a nucleus; $\vec{V}_{20}^{(L)}$ is the
nucleus speed in the $L$-system before the collision with a neutron;
$\vec{V}_{2}^{(L)}$ is the nucleus speed in the $L$-system after the collision
with a neutron; $\vec{V}_{10}^{(L')}$ is the neutron speed in the $L'$-system
before the collision with the nucleus; $\vec{V}_{1}^{(L')}$ is the neutron
speed in the $L'$-system after the collision with a nucleus;
$\vec{V}_{20}^{(L')}$ is the nucleus speed in the $L'$-system before the
collision with a neutron; $\vec{V}_{2}^{(L')}$ is the nucleus speed in the
$L'$-system after the collision with a neutron; $\vec{V}_{c}^{(L')}$ is the
speed of the center of mass in the $L'$-system.

The radius vectors of a point in the laboratory coordinate systems $L$ and $L'$
are connected by the following equation:

\begin{equation}
\vec{r}^{(L)} = \vec{r}_{O'}^{(L)} + \vec{r}^{(L')}
\label{eq01}
\end{equation}
	
Thus, in accordance with our aim, we can write that

\begin{equation}
\vec{V}_{10}^{(L)} = \frac{d \vec{r}_{1}^{(L)}}{dt} \neq 0
 ~\text{and} ~ 
\vec{V}_{20}^{(L)} = \frac{d \vec{r}_{2}^{(L)}}{dt} \neq 0
\label{eq02}
\end{equation}

and choose the $L'$-system so that the nucleus rest in it before the collision:

\begin{equation}
\vec{V}_{20}^{(L')} = 0
\label{eq03}
\end{equation}

The relationship between the coordinates of $m_1$ and $m_2$ in the
$L$-system and the $L'$-system is given by (1), and between the speeds -- by
the following expressions (the Galilean law):

\begin{equation}
\vec{V}_{1}^{(L')} = \vec{V}_{1}^{(L)} - \vec{V}_{20}^{(L)}
~\text{and} ~ 
\vec{V}_{2}^{(L')} = \vec{V}_{2}^{(L)} - \vec{V}_{20}^{(L)}
\label{eq04}
\end{equation}
  
It is convenient to solve the problem of the two particles collision in the
coordinate system associated with the mass center -- the $C$-system.

On the basis of the momentum conservation law, for the two colliding particles
in the $C$-system we obtain:

\begin{equation}
\vec{P}_{10}^{(C)} + \vec{P}_{20}^{(C)} = \vec{P}_{1}^{(C)} + \vec{P}_{2}^{(C)} = 0
\label{eq05}
\end{equation}
  
From the relations~(\ref{eq05}) follows:

\begin{equation}
m_1 \cdot \vec{V}_{10}^{(C)} = - m_2 \cdot \vec{V}_{20}^{(C)}
~\text{and}~
m_1 \cdot \vec{V}_{1}^{(C)} = - m_2 \cdot \vec{V}_{2}^{(C)}
\label{eq06}
\end{equation}

From~(\ref{eq06}) we obtain the speed moduli:

\begin{equation}
\left \vert \vec{V}_{10}^{(C)} \right \vert = V_{10}^{(C)} =
\frac{m_2}{m_1} \left \vert \vec{V}_{20}^{(C)} \right \vert =
\frac{m_2}{m_1} V_{20}^{(C)}
~\text{and}~
V_1^{(C)} = \frac{m_2}{m_1} V_{2}^{(C)}
\label{eq07}
\end{equation}

Introducing the mass number for the neutron and the nucleus $A_n = 1$ and
$A_N = A$, respectively, and assuming that $m_1 = m_n \approx A_n = 1$ and
$m_2 = m_N \approx A_N = A$, for the Eqs.~(\ref{eq06}) and~(\ref{eq07}) we
obtain the following expressions:

\begin{equation}
\vec{V}_{10}^{(C)} = - A \cdot \vec{V}_{20}^{(C)}
~\text{and}~
\vec{V}_{1}^{(C)} = - A \cdot \vec{V}_{2}^{(C)}
\label{eq08}
\end{equation}

\noindent and

\begin{equation}
V_{10}^{(C)} = + A \cdot V_{20}^{(C)}
~\text{and}~
V_{1}^{(C)} = + A \cdot V_{2}^{(C)}
\label{eq09}
\end{equation}

From the kinetic energy conservation law in the C-system we obtain:

\begin{equation}
\frac{\left(V_{10}^{(C)}\right)^2}{2} + \frac{A \cdot \left(V_{20}^{(C)}\right)^2}{2} =
\frac{\left(V_{1}^{(C)}\right)^2}{2} + \frac{A \cdot \left(V_{2}^{(C)}\right)^2}{2} .
\label{eq10}
\end{equation}

Hence, in view of~(\ref{eq09}) and~(\ref{eq10}) we obtain the following
relation:

\begin{equation}
\left(V_{10}^{(C)}\right)^2 \cdot \left( 1 + \frac{1}{A} \right) =
\left(V_{1}^{(C)}\right)^2 \cdot \left( 1 + \frac{1}{A} \right) .
\label{eq11}
\end{equation}

From~(\ref{eq11}) it follows that:

\begin{equation}
V_{10}^{(C)} = V_{1}^{(C)} .
\label{eq12}
\end{equation}

Also, from the kinetic energy conservation law~(\ref{eq10}), taking into
account the relation~(\ref{eq12}) it follows:

\begin{equation}
V_{20}^{(C)} = V_{2}^{(C)} .
\label{eq13}
\end{equation}

\begin{figure}
\begin{center}
\includegraphics[width=9cm]{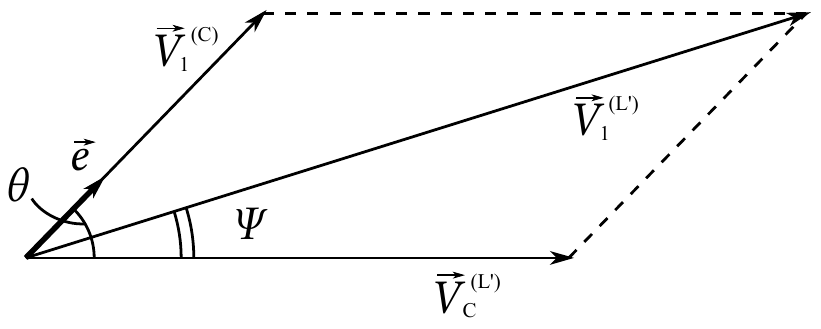}
\end{center}
\caption{Fig.2. Parallelogram of velocities after the collision in the $L'$ coordinate system.}
\label{fig02}
\end{figure}

Thus we can see that in the $C$-system of the closed mechanical system
consisting of the neutron and the nucleus, the neutron and the nucleus move
along a straight line connecting their centers (towards each other before the
collision and in the opposite directions after collision) (see~(\ref{eq06})).
They also decline at certain angle $\theta$ while maintaining the absolute
values of their velocities (see~(\ref{eq12}) and~(\ref{eq13}), 
Fig.~\ref{fig02}).

The center of mass coordinate of the neutron and the moderating
medium nucleus ($\vec{r}_{c}^{(L')}$ is the radius vector of the center of mass
in the $L'$ laboratory coordinate system) may be given as:

\begin{equation}
\vec{r}_{c}^{(L')} = \left(1 \cdot \vec{r}_{1}^{(L')} + 
A \cdot \vec{r}_{2}^{(L')} \right) \cdot \frac{1}{A+1}
\label{eq14}
\end{equation}

\noindent
where $\vec{r}_{1}^{(L')}$ is the radius vector of the neutron in the
$L'$-system; $\vec{r}_{2}^{(L')}$ is the radius vector of the nucleus; $A$ is
the mass number of the nucleus; $1$ is the mass number of a neutron.

Given the fact that in the $L'$-system the nucleus velocity before the
collision $\vec{V}_{20}^{(L')} = 0$, the velocity of the mass center of a
closed system of two particles (the neutron and the nucleus) in the $L'$
coordinate system is:

\begin{equation}
\vec{V}_{C}^{(L')} = \frac{1}{A+1} \cdot \vec{V}_{10}^{(L')} .
\label{eq15}
\end{equation}

By virtue of the law of the total momentum conservation, the velocity of the
inertia center in the $L'$-system will not change after the collision.
Therefore the indices corresponding to the velocity values of the inertia
center before the interaction and after the interaction may be omitted.

Since the mass center system ($C$-system) moves relative to the laboratory
system with the speed of the center of mass in the $L'$-system, for the
velocity of the neutron in the $C$-system before the interaction we have:

\begin{equation}
\vec{V}_{10}^{(C)} = \vec{V}_{10}^{(L')} - \vec{V}_{C}^{(L')} ,
\label{eq16}
\end{equation}

\noindent
then we substitute the expression~(\ref{eq15}) into this expression and obtain:

\begin{equation}
\vec{V}_{10}^{(C)} = \vec{V}_{10}^{(L')} - 
\frac{1}{A+1} \cdot \vec{V}_{10}^{(L')} = 
\frac{A}{A+1} \cdot \vec{V}_{10}^{(L')} .
\label{eq17}
\end{equation}

Using the Eq.~(\ref{eq05}) with regard for the expression~(\ref{eq17}), we find
the speed of the nucleus in the $C$-system before the collision:

\begin{equation}
\vec{V}_{10}^{(C)} = - \frac{1}{A+1} \cdot \vec{V}_{10}^{(L')} .
\label{eq18}
\end{equation}

According to~(\ref{eq12}), (\ref{eq13}) and using the expressions~(\ref{eq17})
and~(\ref{eq18}) we obtain:

\begin{equation}
V_{1}^{(C)} = V_{10}^{(C)} = \frac{A}{A+1} \cdot V_{10}^{(L')}
~\text{and}~
V_{2}^{(C)} = V_{20}^{(C)} = \frac{1}{A+1} \cdot V_{10}^{(L')} .
\label{eq19}
\end{equation}

These results should be adapted to the $L'$ coordinate system. The neutron
velocity in the $L'$ laboratory coordinate system after the collision is:

\begin{equation}
\vec{V}_{1}^{(L')} = \vec{V}_{1}^{(C)} + \vec{V}_{C}^{(L')} ,
\label{eq20}
\end{equation}

Taking into account~(\ref{eq15}), for~(\ref{eq20}) we obtain:

\begin{equation}
\vec{V}_{1}^{(L')} = \vec{V}_{1}^{(C)} + \frac{1}{A+1} \cdot \vec{V}_{10}^{(L')} .
\label{eq21}
\end{equation}

The neutron velocity in the $L'$-system after the collision is directed at an
angle $\Psi$ to the original direction (Fig.~\ref{fig02}).

From the parallelogram of velocities shown in Fig.~\ref{fig02} we find the
squared modulus of the neutron velocity in the $L'$-system after the collision:

\begin{align}
\left( \vec{V}_{1}^{(L')} \right)^2 & = 
\left( V_{10}^{(L')} \cdot \frac{A}{A+1} \right)^2 +
\left( V_{10}^{(L')} \cdot \frac{1}{A+1} \right)^2 + 
\frac{ 2 \cdot A \cdot \left( V_{10}^{(L')} \right)^2}{(A+1)^2} \cdot \cos {\theta} = \nonumber \\
& = \frac{ \left( V_{10}^{(L')} \right)^2 \cdot (A^2 + 2A \cos{\theta} + 1)}{(A+1)^2} ,
\label{eq22}
\end{align}

\noindent
where $\theta$ is the angle of the neutron escape in the $C$-system
(Fig.~\ref{fig02}).

From~(\ref{eq22}) it is possible to find the ratio of the squares of the
velocities before and after neutron interaction with the nucleus in the $L'$
laboratory coordinate system, which is also equal to the ratio of the kinetic
energies of the neutron before and after the interaction:

\begin{equation}
\frac{\left( \vec{V}_{1}^{(L')} \right)^2}{\left( \vec{V}_{10}^{(L')} \right)^2} =
\frac{E_{2}^{(L')}}{E_{1}^{(L')}} = 
\frac{A^2 + 2A \cos{\theta} + 1}{(A+1)^2} ,
\label{eq23}
\end{equation}

\noindent
where $E_1$ and $E_2$ are the kinetic energies of the neutron before and after
the collision in the $L'$-system respectively.

Let us introduce the parameter

\begin{equation}
\alpha = \left( \frac{A-1}{A+1} \right)^2 ,
\label{eq24}
\end{equation}

\noindent
then~(\ref{eq01}) may be given in the following widely-known
form~\cite{BartolomeyBat1989,Shirokov1998}:

\begin{equation}
\frac{E_{2}^{(L')}}{E_{1}^{(L')}} = 
\frac{1}{2} \left[ (1+\alpha) + (1-\alpha) \cos{\theta} \right] .
\label{eq25}
\end{equation}

The maximum value of $E_{2}^{(L')} / E_{1}^{(L')}$ corresponds to the value
$\theta = 0$. This the minimum loss of energy, i.e. the neutron does not lose
its energy in a collision. When $\theta = \pi$ (central collision) 
$E_{2}^{(L')} / E_{1}^{(L')} = \alpha$. The minimum value to which the neutron
energy may reduce as a result of a single elastic scattering is 
$\alpha E_{1}^{(L')}$.

The maximum relative energy loss in a single scattering is

\begin{equation}
\frac{E_{1}^{(L')} - E_{2}^{(L')}}{E_{1}^{(L')}} = 1 - \alpha ,
\label{eq26}
\end{equation}

\noindent
and the maximum possible absolute energy loss is

\begin{equation}
E_{1}^{(L')} (1 - \alpha) .
\label{eq27}
\end{equation}

\noindent
$\alpha = 0$ for hydrogen, i.e. the neutron can lose all its kinetic energy in
a single collision event with a hydrogen nucleus. If we expand $\alpha$ in
powers of $\frac{1}{A}$, we obtain:

\begin{equation}
\alpha = 1 - \frac{4}{A} + \frac{8}{A^2} - \frac{12}{A^3} + \ldots
\label{eq28}
\end{equation}

At $A>50$

\begin{equation}
\alpha \approx 1 - \frac{4}{A}, ~~~\text{a}~~~
1 - \alpha \approx \frac{4}{A} .
\label{eq29}
\end{equation}

It is evident that the smaller is $A$, the greater is the relative energy loss.
Thus the moderation is more effective for the nuclei with small $A$. When 
$A>200$, $(1-\alpha)<2\%$. It means that uranium may not be considered a
moderator in nuclear reactors.

Next we use the Eq.~(\ref{eq04}) connecting the neutron velocities in the $L'$
and $L$ coordinate systems, and the ratio~(\ref{eq25}), to give the
relation~(\ref{eq23}) in the following form:

\begin{equation}
\frac{\left( \vec{V}_{1}^{(L)} - \vec{V}_{20}^{(L)} \right)^2}
{\left( \vec{V}_{10}^{(L)} - \vec{V}_{20}^{(L)} \right)^2} = 
\frac{A^2 + 2A \cos{\theta} + 1}{(A+1)^2} = 
\frac{1}{2} \left[ (1+\alpha) + (1-\alpha) \cos{\theta} \right] .
\label{eq30}
\end{equation}

From~(\ref{eq30}) after some algebraic manipulations it is possible to find the
ratio of the squares of the neutron velocity before and after the interaction
with the nucleus in the $L$-system. It is also equal to the ratio of the
kinetic energy of the neutron before and after the interaction.

\begin{align}
\frac{\left( \vec{V}_{1}^{(L)} \right)^2}{\left( \vec{V}_{10}^{(L)} \right)^2}
& = \frac{E_{1}^{(L)}}{E_{10}^{(L)}} = 
\frac{1}{2} \left[ (1+\alpha) + (1-\alpha) \cos{\theta} \right]
\left[ 1 - 2\frac{V_{10}^{(L)} V_{20}^{(L)} \cos{\beta}}{\left( V_{10}^{(L)} \right)^2} + 
\frac{\left( V_{20}^{(L)} \right)^2}{\left( V_{10}^{(L)} \right)^2} \right] + \nonumber \\
& + 2 \frac{\left( \vec{V}_{1}^{(L)}, \vec{V}_{20}^{(L)} \right)}{\left( V_{10}^{(L)} \right)^2} -
\frac{\left( V_{20}^{(L)} \right)^2}{\left( V_{10}^{(L)} \right)^2} = 
\frac{1}{2} \left[ (1+\alpha) + (1-\alpha) \cos{\theta} \right] - \nonumber \\
& - \left[ (1+\alpha) + (1-\alpha) \cos{\theta} \right] 
\frac{V_{10}^{(L)} V_{20}^{(L)} \cos{\beta}}{\left( V_{10}^{(L)} \right)^2} +
2 \frac{V_{1}^{(L)} V_{20}^{(L)} \cos{\gamma}}{\left( V_{10}^{(L)} \right)^2} - \nonumber \\
& - \left \lbrace 1 - \frac{1}{2} \left[ (1+\alpha) + (1-\alpha) \cos{\theta} \right] \right \rbrace
\frac{1 \cdot E_{20}^{(L)}}{A \cdot E_{10}^{(L)}} ,
\label{eq31}
\end{align}

\noindent
where $\cos{\beta}$ is the cosine of the angle between the vectors 
$\vec{V}_{10}^{(L)}$ and $\vec{V}_{20}^{(L)}$ which is given by the scalar
product of these vectors $\left( \vec{V}_{10}^{(L)},\vec{V}_{20}^{(L)} \right)$
as follows:

\begin{equation}
\cos{\beta} = \frac{\left( \vec{V}_{10}^{(L)}, \vec{V}_{20}^{(L)} \right)}
{\left \vert \vec{V}_{10}^{(L)} \right \vert \left \vert \vec{V}_{20}^{(L)} \right \vert} =
\frac{\left( \vec{V}_{10}^{(L)}, \vec{V}_{20}^{(L)} \right)}
{V_{10}^{(L)} V_{20}^{(L)}} ,
\label{eq32}
\end{equation}

\noindent
$\cos{\gamma}$ is the cosine of the angle between the vectors 
$\vec{V}_{1}^{(L)}$ and $\vec{V}_{20}^{(L)}$ which is given by the scalar
product of these vectors $\left( \vec{V}_{1}^{(L)},\vec{V}_{20}^{(L)} \right)$
as follows:

\begin{equation}
\cos{\gamma} = \frac{\left( \vec{V}_{1}^{(L)}, \vec{V}_{20}^{(L)} \right)}
{\left \vert \vec{V}_{1}^{(L)} \right \vert \left \vert \vec{V}_{20}^{(L)} \right \vert} =
\frac{\left( \vec{V}_{1}^{(L)}, \vec{V}_{20}^{(L)} \right)}
{V_{1}^{(L)} V_{20}^{(L)}} ,
\label{eq33}
\end{equation}

Let us note that from~(\ref{eq31}) it follows that, since the $\cos{\beta}$ and
$\cos{\gamma}$ may take both the positive and negative values, the energy of a
scattered neutron may be both smaller and greater than its initial energy. This
is because some part of the kinetic energy of the nucleus may be transferred
to neutron during the scattering.


As will be shown below, for our purposes it is sufficient to limit oneself to
Eq.~(\ref{eq39}) and not perform the rest of the transformations of the
Eq.~(\ref{eq31}) and related Eqs.~(\ref{eq32}) and~(\ref{eq33}), since they
contain the cosines of the angles between the neutron and nucleus velocity
vectors, which also require the adaptation to the $L$-system.

\section{The neutron scattering law taking into account the thermal motion of the moderating medium nuclei}
\label{sec03}

According to the kinematics of the neutron scattering by a nucleus, shown in
section~\ref{sec02}, the probability for a neutron with kinetic energy
$E_{10}^{(L)}$ before scattering in the $L$-system to possess the kinetic
energy in the range from $E_{1}^{(L)}$ to $E_{1}^{(L)} + dE_{1}^{(L)}$ after
the scattering may be written as follows:

\begin{align}
P \left( E_{1}^{(L)} \right) dE_{1}^{(L)} & = 
P \left( \theta, \beta, \gamma, E_{N}^{(L)} \right) d\theta d\beta d\gamma dE_{N}^{(L)} =
\nonumber \\
& = P ( \theta ) d\theta \cdot P( \beta) d\beta \cdot P(\gamma) d\gamma \cdot 
P\left(E_{N}^{(L)} \right) dE_{N}^{(L)} .
\label{eq34}
\end{align}

From the quantum mechanical scattering theory it is known
(e.g.~\cite{Feinberg1978,Levich1971}) that if the de Broglie wavelength is much
larger than the size of the nucleus, the neutron scattering by a potential well
of the nucleus must be spherically symmetric in the center of inertia
coordinate system ($C$-system), i.e. isotropic. It is useful to estimate the
threshold neutron energy $E_B$ above which the neutron scattering is not
spherically symmetric. According to \cite{Feinberg1978}, this can be done using
the quasiclassical approximation in the quantum-mechanical problem of the
neutron scattering by a potential well of the nucleus. Let $R$ be the effective
radius of the nucleus, $\rho$ -- the impact parameter of the incident neutron,
$v$ -- its speed. As in the classical representation of the orbital angular
momentum of the neutron $\rho v = l \hbar (l = 0,1,2,\ldots)$, the scattering
should take place at $R v \geqslant \rho v = l \hbar$. The scattering is
spherically symmetric if $l = 0$, and only if $l \geqslant 1$, the scattering
is the anisotropic. Therefore, for the anisotropic scattering the condition 
$R v \geqslant \hbar$ must be satisfied, which implies that 
$v _B \geqslant \hbar / R$. Since $E_B = m v^2 /2$, where $m$ is the neutron
mass, and the nucleus radius may be expressed as a well known expression
$R \approx r_0 A^{1/3}$, where $r_0 = 1.2 \cdot 10^{-13}$cm, $A$ is the mass
number of the scattering nucleus, the neutron energy threshold is

\begin{equation}
E_B = \frac{\hbar ^2}{2 m r_0 ^2} \frac{1}{A^{2/3}}
\sim \frac{10 ~MeV}{A^{2/3}} .
\label{eq35}
\end{equation}

Thus, according to~(\ref{eq35}) for hydrogen as moderator, almost the entire
range of fission spectrum (0 - 10~MeV) is the range of the spherically
symmetric scattering of neutrons in the $C$-system. For carbon ($A$ = 12) we
obtain that $E_B \sim 2~MeV$. Because the average energy of the fission neutron
spectrum is $\sim 2~MeV$, it gives a reason to believe that the neutron
moderation by the light nuclei, the scattering is spherically symmetric in the
$C$-system.

These estimates are confirmed by the experimental data, 
e.g.~\cite{BartolomeyBat1989}, the neutron scattering is spherically symmetric
in the center of mass coordinate system (isotropic) up to the neutron energy
$\sim 10^5 ~eV$.

Thus, since the scattering of neutrons in the center of mass coordinate system
is spherically symmetric (isotropic), then for $P(\theta ) d\theta$ we obtain:

\begin{equation}
P(\theta ) d\theta = \int \limits _{0} ^{2 \pi} 
\left[ P(\theta , \varphi ) d\theta \right] d \varphi = 
\int \limits _{0} ^{2\pi} \frac{r \sin{\theta} d\varphi \cdot r d\theta}{4 \pi r^2} =
\frac{\sin{\theta} d\theta}{4 \pi} \int \limits _{0} ^{2 \pi} d\varphi = 
\frac{1}{2} \sin{\theta} d\theta ,
\label{eq36}
\end{equation}

\noindent
where $\varphi$ is the azimuth angle of the usual spherical coordinates
$r,\theta, \varphi$ introduced in the center of mass coordinate system.

Since the thermal motion of the moderating medium nuclei is chaotic, and the
neutron source is isotropic, the distribution of the velocity vector directions
in the space for the neutrons after the collision is equiprobable over the
$\beta$ and $\gamma$ angles, i.e. also spherically symmetric (isotropic). So by
analogy we obtain:

\begin{equation}
P(\beta ) d\beta = \frac{1}{2} \sin{\beta} d\beta ,
\label{eq37}
\end{equation}

\begin{equation}
P(\gamma ) d\gamma = \frac{1}{2} \sin{\gamma} d\gamma ,
\label{eq38}
\end{equation}

By averaging the neutron kinetic energy after scattering given by the
expression~(\ref{eq31}) over the spherically symmetric distribution of the
moderator nuclei thermal motion and isotropic neutron source, we obtain the
following expression:

\begin{align}
\bar{E}_1^{(L)} & = \int \limits _{0} ^{\pi} \int \limits _{0} ^{\pi} 
E_1 ^{(L)} \cdot P(\beta) d\beta \cdot P(\gamma) d\gamma = \nonumber \\
& = \bar{E}_{10}^{(L)} \left \lbrace 
\frac{1}{2} \left[ (1+\alpha) + (1-\alpha) \cos{\theta} \right] -
\left[ 1 - \frac{1}{2} \left[ (1+\alpha) + (1-\alpha) \cos{\theta} \right] 
\frac{E_N^{(L)}}{A \cdot \bar{E}_{10}^{(L)}} \right] \right \rbrace .
\label{eq39}
\end{align}

Here $\bar{E}_{10}^{(L)}$ is the neutron energy averaged over the neutron
momenta directions for the isotropic neutron source (coincides with the neutron
energy $\bar{E}_{10}^{(L)} = E_{10}^{(L)}$), $E_N^{(L)}$ is defined by the
Maxwell distribution~\cite{Levich1969} for the moderator nuclei, which depends
on the moderating medium temperature:

\begin{equation}
P \left( E_N^{(L)} \right) dE_N^{(L)} = \frac{2}{\sqrt{\pi (kT)^3}} 
e^{-\frac{E_N^{(L)}}{kT}} \sqrt{E_N^{(L)}} dE_N^{(L)} .
\label{eq40}
\end{equation}

Let us average the expression~(\ref{eq39}) over the Maxwellian distribution of
the thermal motion of the moderating medium nuclei~(\ref{eq40}), considering
that $\bar{E}_{10}^{(L)} = E_{10}^{(L)}$. If we use a well-known result
$\bar{E}_N^{(L)} = \int \limits _{0} ^{\infty} E_N^{(L)} 
P_M \left( E_N^{(L)} \right) dE_N^{(L)} = \frac{3}{2} kT$~\cite{Levich1969}, we
obtain the following expression:

\begin{align}
\bar {\bar{E}}_{1}^{(L)} & = \int \limits _{0} ^{\infty} \bar{E}_{1}^{(L)}
P_M \left( E_N^{(L)} \right) dE_N^{(L)} = \nonumber \\
& = E_{10}^{(L)} \left \lbrace 
\frac{1}{2} \left[ (1+\alpha) + (1-\alpha) \cos{\theta} \right] -
\left[ 1 - \frac{1}{2} \left[ (1+\alpha) + (1-\alpha) \cos{\theta} \right] 
\frac{\frac{3}{2}kT}{A \cdot E_{10}^{(L)}} \right]
\right \rbrace
\label{eq41}
\end{align}

Since the functional relationship between $\bar {\bar{E}}_{1}^{(L)}$ and
$\theta$ unique, as follows from~(\ref{eq41}), the probability
$P \left( \bar {\bar{E}}_{1}^{(L)} \right) dE_1^{(L)}$ for the neutron with a
kinetic energy $E_{10}^{(L)}$ in the $L$-system before scattering to possess
the kinetic energy in the range from $\bar {\bar{E}}_{1}^{(L)}$ to
$\bar {\bar{E}}_{1}^{(L)} + dE_1^{(L)}$ after the scattering on the chaotically
moving moderator nuclei is determined by the $P(\theta) d\theta$
distribution~(\ref{eq36}). Therefore we obtain the following relation (here we
omit the symbols of averaging and the laboratory coordinate system $L$ for
simplicity, i.e. 
$P \left( \bar {\bar{E}}_{1}^{(L)} \right) dE_1^{(L)} = P(E_1) dE_1$):

\begin{align}
P(E_1) dE_1 & = P(\theta) d\theta = 
P(\theta) \left \vert \frac{d\theta}{dE_1} \right \vert dE_1 = \nonumber \\
& = \frac{1}{2} \sin{\theta} \left \vert \frac{1}{E_{10}^{(L)} \left[ 
\frac{1}{2} (1 - \alpha) \sin{\theta} + \frac{1}{2} (1-\alpha) \sin{\theta} 
\frac{\frac{3}{2}kT}{A \cdot E_{10}^{(L)}} \right]} \right \vert dE_1 = \nonumber \\
& = \frac{dE_1}{\left[ E_{10}^{(L)} + \frac{1}{A} \cdot \frac{3}{2} kT \right]
(1-\alpha)}
\label{eq42}
\end{align}

Thus, we obtained the neutron scattering law, which takes into account the
thermal motion of the moderating medium nuclei:

\begin{equation}
\begin{cases}
P(E_1) dE_1 = 
\frac{dE_1}{\left[ E_{10}^{(L)} + \frac{1}{A} \frac{3}{2} kT \right] 
(1-\alpha)} & \text{when}~~ \alpha \left( E_{10}^{(L)} + \frac{1}{A} \frac{3}{2} kT \right)
 \leqslant E_1 \leqslant \left( E_{10}^{(L)} + \frac{1}{A} \frac{3}{2} kT \right)\\
P(E_1) = 0 & \text{when}~~ E_1 \leqslant 
\alpha \left( E_{10}^{(L)} + \frac{1}{A} \frac{3}{2} kT \right)
~\text{,}~ E_1 \geqslant 
\left( E_{10}^{(L)} + \frac{1}{A} \frac{3}{2} kT \right)
\end{cases}
\label{eq43}
\end{equation}

In conclusion of the section, let us emphasize that the new scattering
law~(\ref{eq43}) is written for the averaged neutron energy after scattering
$E_1$. The averaging of the neutron energy is performed over the thermal
(chaotic) motion of the moderating medium nuclei and the neutron source
isotropy.


It should also be noted that although (as mentioned in section~\ref{sec02}
above) the relation~(\ref{eq31}) implies that the individual neutrons may
receive an additional energy as a result of scattering, the scattering law for
an isotropic neutron source~(\ref{eq43}), averaged over the thermal motion of
the moderating medium nuclei and neutron source isotropy yields the energy of a
group of neutrons after scattering almost always smaller than the averaged
energy of a group of neutrons before scattering. It means that this is
actually the neutron moderation law, but now taking into account the thermal
motion of the moderating medium nuclei and the isotropy of the neutron source.


Let us also mention some "tricks" or "moves" it was necessary to take in order
to obtain the neutron moderation law in the form of~(\ref{eq43}), since the
previous attempts failed without them (e.g. an attempt by
Galanin~\cite{Galanin1971}):
\begin{itemize}
\item the first move is the transition from the $L$ laboratory coordinate
system to the $L'$ laboratory coordinate system, in which the moderator nucleus
rests (section~\ref{sec02}). Thus the kinematics of the standard moderation
theory leading to the expression~(\ref{eq30}) may be applied;
\item the second move is the inverse transition from the $L'$ coordinate system
to the $L$-system transforming~(\ref{eq30}) into~(\ref{eq31});
\item the third move is the averaging of the expression~(\ref{eq31}) over the
thermal motion of the nuclei, which gives~(\ref{eq41});
\item the fourth move is the derivation of the scattering law~(\ref{eq43}).
\end{itemize}

As we shall see below, the scattering law~(\ref{eq43}) lets one derive the
expressions for the flux and energy spectra of the moderating neutrons in a
variety of media, taking into account the temperature of the medium.

In order to take the neutron scattering anisotropy into account, a transport
macroscopic cross-section is introduced in the nuclear reactor physics~\cite{Feinberg1978,BartolomeyBat1989,Shirokov1998}:

\begin{equation}
\Sigma _{tr} = \Sigma _{a} + \Sigma _{s} (1 - \bar{\mu}) ,
\label{eq44}
\end{equation}

\noindent
where $\Sigma _{a}$ and $\Sigma _{s}$ are the macroscopic absorption
cross-section and the macroscopic scattering cross-section of neutrons
respectively, $\bar{\mu}$ is the average cosine of the scattering angle. If 
$\Sigma _{a} \ll \Sigma _{s}$, then

\begin{equation}
\Sigma _{tr} \approx \Sigma _{s} (1 - \bar{\mu}) .
\label{eq45}
\end{equation}

For small deviations from spherical symmetry typical for the reactor neutrons,
i.e. for the scattering anisotropy, which may be observed during the scattering
of the high-energy neutrons by the moderators with heavy nuclei, the scattering
law, according to~\cite{BartolomeyBat1989}, may be given as follows:

\begin{equation}
\begin{cases}
P(E_1) dE_1 = 
\frac{dE_1}{\left[ E_{10}^{(L)} + \frac{1}{A} \cdot \frac{3}{2} kT \right] 
(1-\alpha)} \left[ 1 + 3 \bar{\mu} - \frac{6}{(1-\alpha)} \bar{\mu} 
\frac{E_1 - E_{10}^{(L)}}{E_1} \right] & 
\text{when}~~ \alpha E_{10}^{(L)} \leqslant E_1 \leqslant E_{10}^{(L)}\\
P(E_1) = 0 & \text{when}~~ E_1 \leqslant \alpha E_{10}^{(L)} 
~~\text{and}~~ E_1 \geqslant E_{10}^{(L)}
\end{cases}
\label{eq46}
\end{equation}

\noindent where
\begin{equation}
\bar{\mu} = 0.07 \cdot A^{2/3} E ~~~\text{when}~~ l=1 
~~~\text{(see the beginning of this section)} .
\label{eq47}
\end{equation}

In contrast to the scattering law for isotropic scattering~(\ref{eq43}), the 
probability $P(E_1)$ depends on the final neutron energy $E_1$ in the
scattering law~(\ref{eq46}).

It is also obvious that at neutron energies tending to zero, the gas model
considered here stops working, and it is necessary to develop a moderation
theory taking into account the interaction between the nuclei (ions or
molecules) of a moderator.

\section{Neutron moderation in hydrogen media without absorption}
\label{sec04}

According to the scattering law~(\ref{eq43}), the neutron moderation law in the
non-absorbing hydrogen media ($\alpha = 0$ and $A = 1$) has the following form:

\begin{equation}
P(E_1) dE_1 = 
\frac{dE_1}{\left[ E_{10}^{(L)} + \frac{3}{2} kT \right]} 
\label{eq48}
\end{equation}

Performing the calculations similar to those
in~\cite{Feinberg1978,BartolomeyBat1989,Shirokov1998}, using the moderation
law~(\ref{eq48}) we find the following expression for the moderating neutrons
flux density (here the new notation is introduced $E = E_{10}^{(L)}$ and
$dE = dE_1$ for the sake of simplification):

\begin{equation}
\Phi(E) = \frac{\int \limits _{E} ^{\infty} Q(E') dE'}
{\left[ E + \frac{3}{2} kT \right]  \Sigma _s (E)} +
\frac{Q(E)}{\Sigma_s(E)} ,
\label{eq49}
\end{equation}

\noindent
where $Q(E')$ is the number of neutrons generated per unit volume per unit time
with the energy $E'$.

Indeed, according to the neutron scattering law~(\ref{eq43}), the energy of
neutrons having an initial energy $E$ will be distributed equiprobably in the
range from $E$ to $0$ after a collision with hydrogen nuclei.

If we divide this energy range from $E$ to $0$ into intervals of size $dE$,
the number of neutrons scattered in the interval $dE$ as a result of scattering
collisions in the range $dE$ in the vicinity of the energy $E$ will be:

\begin{equation}
\Phi(E) \Sigma_s (E) dE = \frac{dE}{E + \frac{3}{2} kT} = 
F(E) dE \frac{dE}{E + \frac{3}{2} kT} ,
\label{eq50}
\end{equation}

\noindent
where $F(E)$ denotes the number of the acts of neutron scattering with
energy $E$ per unit volume per unit time.

The total number of neutrons scattered into the energy range $dE$ as a result
of all collisions of this type in the energy range from the initial neutron
energy $E_{10}$ to $E$ is

\begin{equation}
dE \int \limits_{E} ^{E_{10}} \frac{F(E') dE'}{E' + \frac{3}{2} kT} .
\label{eq51}
\end{equation}

It is assumed here that the neutrons had been scattered into the interval $dE$
in the vicinity of the energy $E$, and only after that they were scattered
into the interval $dE$. However, since a single collision with a hydrogen
nucleus can reduce the neutron energy from its initial value $E_{10}$ to $0$,
some neutrons are apparently scattered into the interval $dE$ as a result of
their first collision.

A monoenergetic (discrete energy) neutron source can be written mathematically as

\begin{equation}
Q(E) = Q_0 \delta (E - E_{10})
\label{eq52}
\end{equation}

\noindent
where $Q_0 = Q(E_{10})$ is the number of generated neutrons with the energy
$E_{10}$ per unit volume per unit time, $\delta (E - E_{10})$ is the
generalized Dirac function.

As for the case of non-monoenergetic neutron source (continuous spectrum with a
maximum of neutron energy of $E_{max}$), the total number of neutrons scattered
into an interval of energy $dE$ near the energy $E$ is

\begin{align}
dE \left[ \int \limits_{E} ^{E_{max}} \left( \int \limits_{E} ^{E_{10}} 
\frac{F(E') dE'}{E' + \frac{3}{2} kT} + 
Q(E_{10}) \delta (E - E_{10}) \right) dE_{10} \right] = \nonumber \\
= dE \left[ \int \limits_{E} ^{E_{max}} \left( \int \limits_{E} ^{E_{10}} 
\frac{F(E') dE'}{E' + \frac{3}{2} kT} \right) dE_{10} + 
Q(E) \right] .
\label{eq53}
\end{align}

The number of the neutrons leaving interval of energies $dE$ near $E$ due to
scattering is equal to $F(E) dE$, so the non-stationary neutron balance
equation for the interval of energies $dE$ near $E$ is

\begin{equation}
\frac{\partial n (E,t)}{\partial t} dE =
dE \left[ \int \limits_{E} ^{E_{max}} \left( \int \limits_{E} ^{E_{10}} 
\frac{F(E') dE'}{E' + \frac{3}{2} kT} \right) dE_{10} + 
Q(E) \right]  - F(E) dE .
\label{eq54}
\end{equation}

The non-stationary integro-differential equation~(\ref{eq54}) cannot be solved
analytically, and its solutions may only be found numerically. Of course, in
the future we will work on the development of the theory of non-stationary
moderation process, but in this paper we will focus on a more simple stationary
case.

The condition of the stationarity of the neutron energy distribution means that
the number of neutrons leaving each elementary energy interval as a result of
the scattering should be equal to the total number of neutrons entering the
same interval. The number of neutrons leaving the interval $dE$ due to
scattering is $F(E)dE$, so the condition of stationarity for interval $dE$ may
be written as

\begin{equation}
F(E) dE =
dE \left[ \int \limits_{E} ^{E_{max}} \left( \int \limits_{E} ^{E_{10}} 
\frac{F(E') dE'}{E' + \frac{3}{2} kT} \right) dE_{10} + 
Q(E) \right] .
\label{eq55}
\end{equation}

Equation~(\ref{eq55}) in the case of a stationary moderation spectrum is
equivalent to the following equation:

\begin{equation}
F(E) dE =
\left[ \int \limits_{E} ^{E_{10}} 
\frac{F(E') dE'}{E' + \frac{3}{2} kT} + Q(E) \right] dE.
\label{eq56}
\end{equation}

According to~\cite{Feinberg1978,BartolomeyBat1989,Shirokov1998}, the integral
equation~(\ref{eq56}) may be reduced to a differential equation with the
following solution:

\begin{equation}
F(E) = \frac{ \int \limits_{E} ^{E_{max}} Q(E') dE'}{\left[ E + \frac{3}{2} kT\right]} + Q(E)  .
\label{eq57}
\end{equation}


Indeed, if we differentiate~(\ref{eq56}) with respect to energy $E$, we obtain

\begin{equation}
\frac{dF(E)}{dE} = - \frac{F(E)}{\left[ E + \frac{3}{2} kT\right]} + \frac{dQ(E)}{dE} .
\label{eq57a}
\end{equation}

By putting the derivatives into left hand side, we obtain a non-homogeneous
differential equation

\begin{equation}
\frac{d \left[ F(E) - Q(E) \right]}{dE} = - \frac{F(E)}{\left[ E + \frac{3}{2} kT\right]} .
\label{eq57b}
\end{equation}

According to the differential equation theory, the general solution of a
non-homogeneous differential equation~(\ref{eq57b}) is a sum of the solution of
the corresponding homogeneous differential equation, which may be obtained
from~(\ref{eq57b}) by zeroing out its right-hand side, and some particular
solution of the non-homogeneous equation~(\ref{eq57b}).

Let us first find a solution of the homogeneous equation of the form

\begin{equation}
\frac{d \left[ F(E) - Q(E) \right]}{dE} = 0 .
\label{eq57c}
\end{equation}

The solution of the Eq.~(\ref{eq57c}) may be easily found, and has the
following form:

\begin{equation}
F(E) = Q(E) + const .
\label{eq57d}
\end{equation}

Now let us find a particular solution of the non-homogeneous
equation~(\ref{eq57b}). For this purpose let us perform the
gradient calibration, i.e. switch from the function $F(E)$
to the function $F'(E)$ related to $F(E)$ in the following way:

\begin{equation}
F(E) = F'(E) + f(E) ,
\label{eq57e}
\end{equation}

\noindent
where $f(E)$ is the calibration function which must fullfill
the calibration equation we shall derive below.

Let us substitute the expression~(\ref{eq57e}) into non-homogeneous
equation~(\ref{eq57b}) end obtain the following:

\begin{equation}
\frac{d F'(E)}{dE} + \frac{df (E)}{dE} - \frac{dQ(E)}{dE} = 
-\frac{F'(E)}{\left[ E + \frac{3}{2} kT\right]} - 
\frac{f(E)}{\left[ E + \frac{3}{2} kT\right]} .
\label{eq57f}
\end{equation}

We impose the following condition for the calibration function
$f(E)$:

\begin{equation}
\frac{d f(E)}{dE} - \frac{dQ(E)}{dE} + 
\frac{f(E)}{\left[ E + \frac{3}{2} kT\right]} = 0 .
\label{eq57g}
\end{equation}

Then from~(\ref{eq57f}) we obtain a system of two equations

\begin{numcases}{}
\frac{d F'(E)}{dE} = -\frac{F'(E)}{\left[ E + \frac{3}{2} kT\right]} , & \label{eq57h} \\
\frac{d\left[ f(E)-Q(E) \right]}{dE} = -\frac{f(E)}{\left[ E + \frac{3}{2} kT\right]} . &
\label{eq57i}
\end{numcases}

Let us find the solution to the equation~(\ref{eq57h}). Its solution has the
form

\begin{equation}
F'(E) = \frac{C_1}{\left[ E + \frac{3}{2} kT\right]} ,
\label{eq57j}
\end{equation}

\noindent
where $C_1$ is an arbitrary constant.

Now let us consider Eq.~(\ref{eq57i}) for $f(E)$. Its form coincides with that
of the non-homogeneous differential equation~(\ref{eq57b}) for $F(E)$. Thus
the particular solution for the Eq.~(\ref{eq57i}) may be given as follows:

\begin{equation}
f(E) = C_2 \cdot F(E) ,
\label{eq57k}
\end{equation}

\noindent
where $C_2$ is an arbitrary constant.

Substituting the solution~(\ref{eq57k}) into~(\ref{eq57e}), we obtain the
following expression for $F(E)$:

\begin{equation}
F(E) = F'(E) + C_2 \cdot F(E) ,
\label{eq57l}
\end{equation}

From~(\ref{eq57l}) for $F(E)$ we obtain:

\begin{equation}
F(E) = F'(E)/ C_3 ,
\label{eq57m}
\end{equation}

\noindent
where $C_3 = (1 - C_2)$ is a non-zero constant.

Further substituting the solution~(\ref{eq57j}) into~(\ref{eq57m}), for the
particular solution of the non-homogeneous differential equation~(\ref{eq57b})
$F(E)$ we have:

\begin{equation}
F(E) = \frac{C_4}{\left[ E + \frac{3}{2} kT\right]} ,
\label{eq57n}
\end{equation}

\noindent
where $C_4 = C_1 / C_3$ is a constant.

Thus, having the solution~(\ref{eq57d}) of the homogeneous differential
equation and the particular solution of the non-homogeneous
equation~(\ref{eq57n}), we can write down the general solution of the
equation~(\ref{eq57b}):

\begin{equation}
F(E) = \frac{C_4}{\left[ E + \frac{3}{2} kT\right]} + const + Q(E) .
\label{eq57o}
\end{equation}

Let us now find the constants in Eq.~(\ref{eq57o}). For this purpose let us
write the equation~(\ref{eq57}) for the energy $E = E_{10}$ and the
monoenergetic neutron souce~(\ref{eq52}) acting at the energy $E = E_{10}$:

\begin{equation}
F(E_{10}) = \frac{Q_0}{\left[ E_{10} + \frac{3}{2} kT\right]} + Q_0 .
\label{eq57p}
\end{equation}

From the general solution~(\ref{eq57o}) for the same conditions we obtain:

\begin{equation}
F(E_{10}) = \frac{C_4}{\left[ E_{10} + \frac{3}{2} kT\right]} + const + Q_0 .
\label{eq57q}
\end{equation}

When comparing the equations~(\ref{eq57p}) and~(\ref{eq57q}) it becomes evident
that $C_4 = Q_0$ and $const = 0$.

Thus, for the monoenergetic neutron source the general solution of the
non-homogeneous differential equation~(\ref{eq57b}) has the form:

\begin{equation}
F(E) = \frac{Q_0}{\left[ E + \frac{3}{2} kT\right]} + Q_0 \delta (E - E_{10}) .
\label{eq57r}
\end{equation}

Generalizing the obtained solution for the non-monochromatic neutron source we
derive:

\begin{equation}
F(E) = \frac{\int \limits _{E} ^{\infty} Q(E') dE'}{\left[ E + \frac{3}{2} kT\right]} + Q(E) .
\label{eq57s}
\end{equation}


From~(\ref{eq57}) for the neutron flux density we obtain

\begin{equation}
\Phi(E) = \frac{ \int \limits_{E} ^{\infty} Q(E') dE'}
{\left[ E + \frac{3}{2} kT\right] \Sigma _s (E)} + \frac{Q(E)}{\Sigma_s(E)} ,
\label{eq58}
\end{equation}

\noindent
where $Q(E)$ is the number of generated neutrons with energy $E$ per unit
volume per unit time.

Thus, we obtain the expression~(\ref{eq49}).

For the monoenergetic neutron source~(\ref{eq52}) the solution~(\ref{eq58})
or~(\ref{eq49}) takes the following form:

\begin{equation}
\Phi(E) = \frac{ Q_0 }{\left[ E + \frac{3}{2} kT\right] \Sigma _s (E)} + 
\frac{Q_0 \delta (E-E_{10})}{\Sigma_s(E)} .
\label{eq59}
\end{equation}

Let us note that a well known Fermi spectrum 
(e.g.~\cite{Feinberg1978,BartolomeyBat1989,Shirokov1998,Stacey2001}) follows
from Eq.~(\ref{eq59}) for all energies less than the energy of the
monoenergetic source $E_{10}$. Still, the known form of the Fermi spectrum does
not include a term with the temperature. It means that a complete matching of
the Eq.~(\ref{eq59}) with the known expression for the Fermi spectrum happens
at $E \gg \frac{3}{2} kT$.

Given that the neutron density $n(E)$ is equal to (e.g.~\cite{Shirokov1998})

\begin{equation}
n(E) = \frac{\Phi (E)}{\sqrt{2E/m_n}} ,
\label{eq60}
\end{equation}

\noindent
and substituting~(\ref{eq59}) into~(\ref{eq60}), we derive the following
expression for the probability density function of the moderating neutrons
distribution over their energies:

\begin{equation}
\rho(E) = \frac{n(E)}{\int \limits_0 ^{\infty} n(E') dE'} = 
\frac{\frac{1}{\sqrt{2E/m_n}} 
\left \lbrace 
\frac{ \int \limits_{E} ^{\infty} Q(E') dE'}
{\left[ E + \frac{3}{2} kT\right] \Sigma _s (E)} + \frac{Q(E)}{\Sigma_s(E)}
\right \rbrace}
{ \int \limits _{0} ^{\infty} \left[
\frac{1}{\sqrt{2E'/m_n}}
\left \lbrace 
\frac{ \int \limits_{E'} ^{\infty} Q(E'') dE''}
{\left[ E' + \frac{3}{2} kT\right] \Sigma _s (E')} + \frac{Q(E')}{\Sigma_s(E')}
\right \rbrace \right] dE'}
\label{eq61}
\end{equation}

For the reactor fissile media $Q(E)$ is determined by the fission spectrum of
the fissile nuclide (or their combination) which, according
to~\cite{Weinberg1961,Stacey2001,EngPhysRef1961,Shirokov1972}, may be given as

\begin{equation}
Q(E) = Q_0 \cdot c \exp {(-aE)} \text{sh} {\sqrt{bE}} ,
\label{eq62}
\end{equation}

\noindent
where $c$, $a$ and $b$ are the constants given in Table~\ref{tab01} below, $E$
is the neutron energy nondimensionalized by 1~MeV, $Q_0$ is the total number of
neutrons generated per unit volume per unit time.

\begin{table}
\begin{center}
\begin{tabular}{|c|c|c|c|c|}
\hline
var & ${}^{235}$U & ${}^{239}$Pu & ${}^{233}$U & ${}^{241}$Pu \\
\hline
a & 1.036  & 1                   & 1.05$\pm$0.03 & 1.0$\pm$0.05 \\
b & 2.29   & 2                   & 2.8$\pm$0.10  & 2.2$\pm$0.05 \\
c & 0.4527 & $\sqrt{2/{\pi e}}$ & 0.46534       & 0.43892 \\
\hline
\end{tabular}

\end{center}
\caption{The constants defining the fission spectrum of the major reactor
         fissile nuclides.}
\label{tab01}
\end{table}

\begin{figure}[htb]
\begin{center}
\includegraphics[width=14cm]{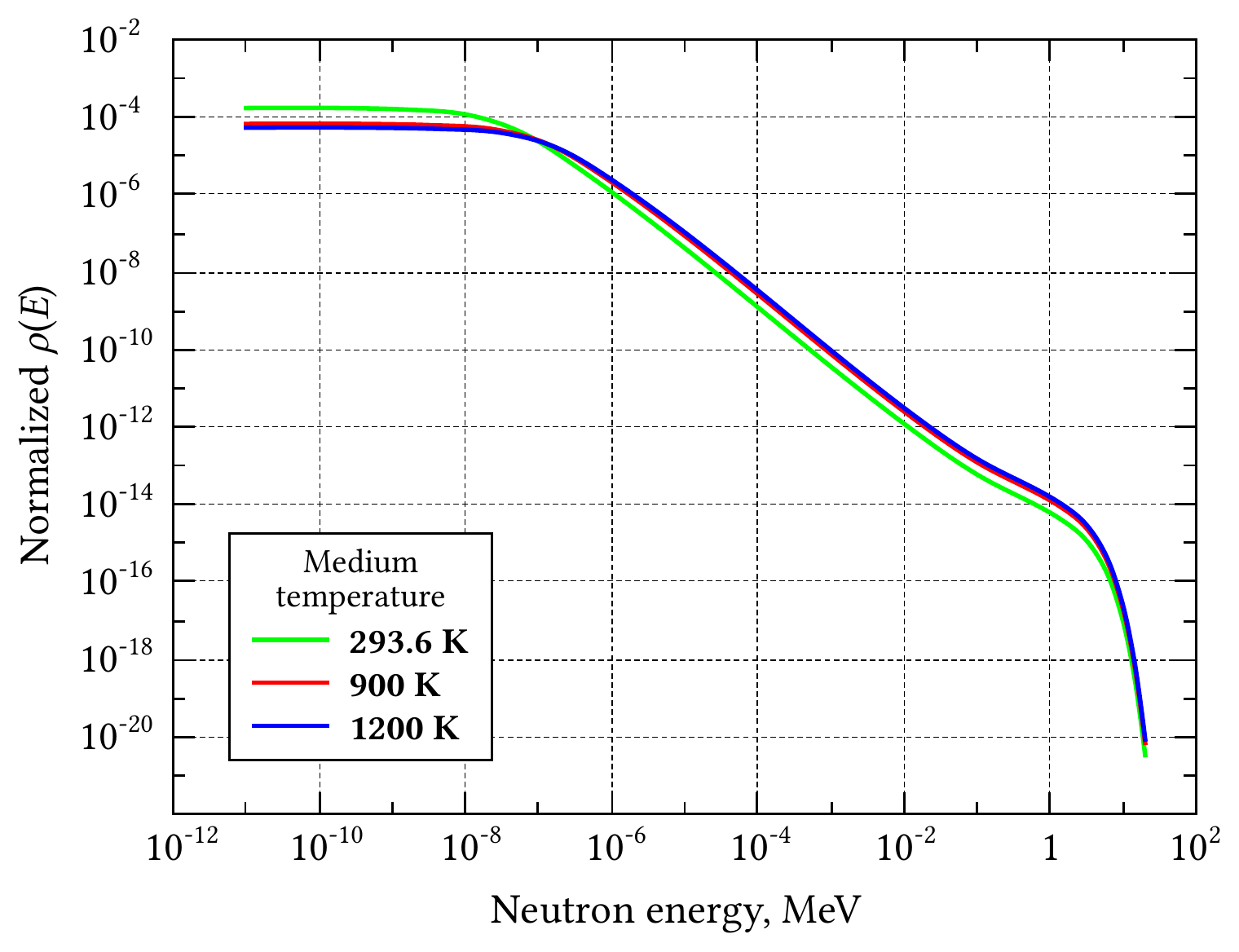}
\end{center}
\caption{The neutron spectrum (0 - 5~MeV) calculated by the 
         expression~(\ref{eq61}) for the source of fission neutrons given by
         the expression~(\ref{eq62}) for ${}^{235}$U, and the moderator
         temperature of 1000~K.}
\label{fig03}
\end{figure}

Fig.~\ref{fig03} shows the energy spectrum of the moderating neutrons
calculated using the expression~(\ref{eq61}) for the fission neutron source
given by expression~(\ref{eq62}) for ${}^{235}$U (Table~\ref{tab01}) at a
temperature of the moderating medium of 1000~K. The macroscopic elastic
scattering cross-section of the neutrons on hydrogen is shown in
Fig.~\ref{fig05} and was taken from the ENDF/B-VII.0 database.

The analysis of the energy spectrum shown in Fig.~\ref{fig03} shows that a
single expression~(\ref{eq61}) describes the energy spectrum of the
moderating neutrons completely and physically correctly, taking into account
the moderating medium temperature.

\begin{figure}[htb]
\begin{center}
\includegraphics[width=9cm]{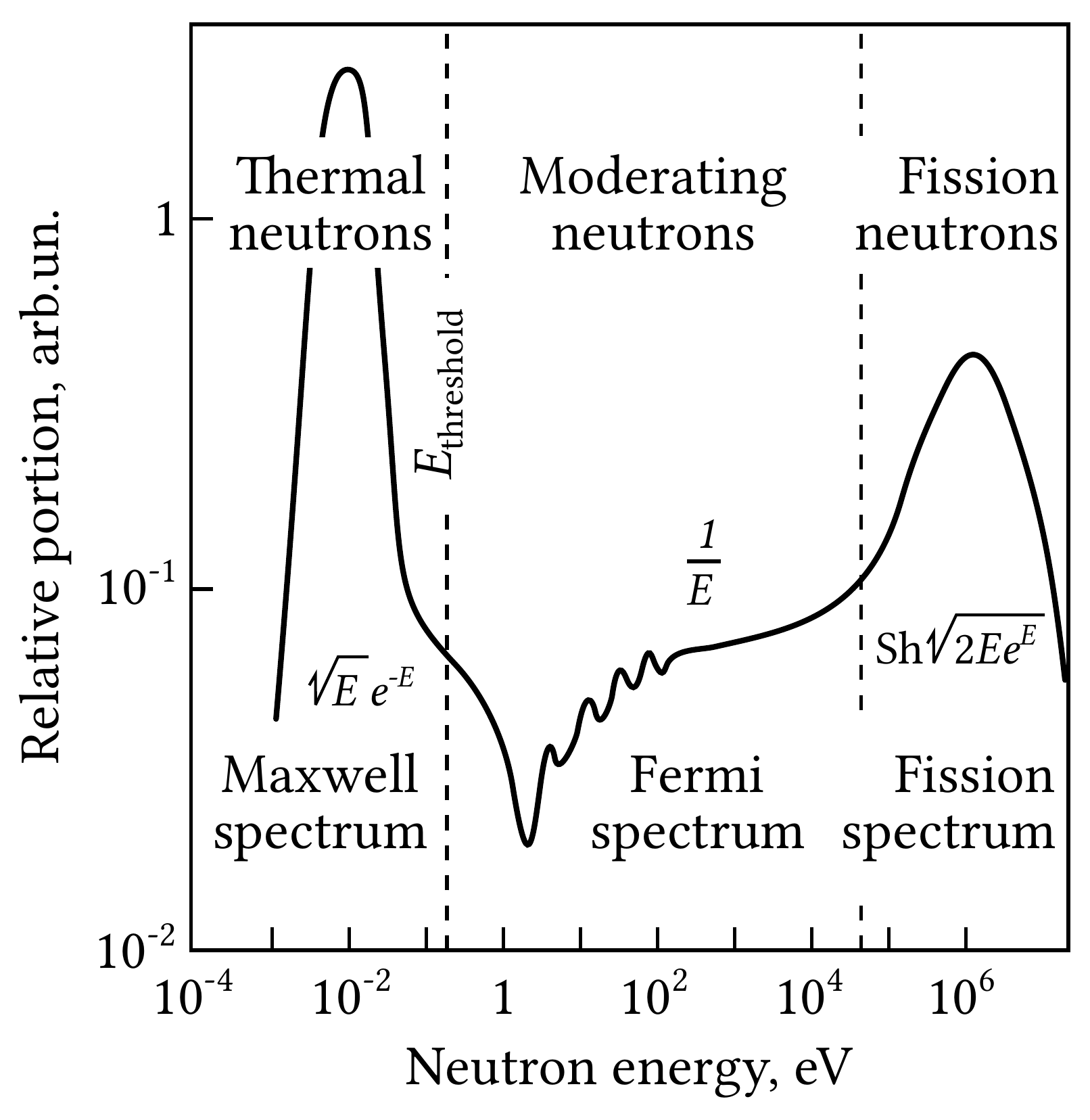}
\end{center}
\caption{Theoretical scheme of a full energy spectrum of moderating
         neutrons~\cite{Vladimirov1986}.}
\label{fig04}
\end{figure}

\begin{figure}
\begin{center}
\includegraphics[width=12cm]{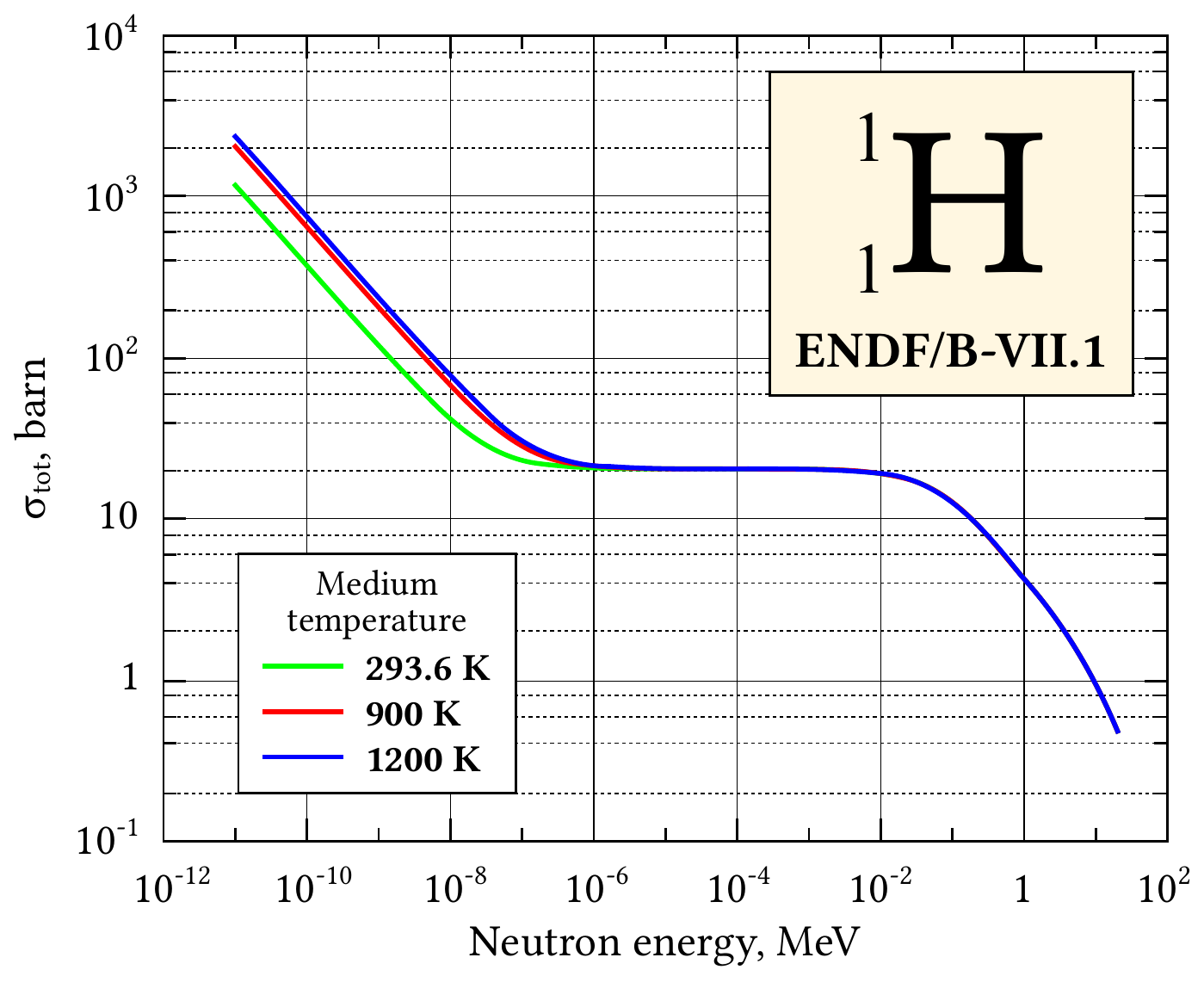}
\end{center}
\caption{Dependence of microscopic cross-section of the neutron elastic
         scattering on hydrogen on the neutron energy (from the ENDF/B-VII.1
         database).}
\label{fig05}
\end{figure}

Let us compare the energy spectrum of the moderating neutrons, given by
expression~(\ref{eq61}), with a scheme of the complete energy spectrum of
moderating neutrons sometimes found in the literature~\cite{Vladimirov1986} and
shown in Fig.~\ref{fig04}. The following not quite correct physical
representation of the nature of the neutron spectrum form may be noticed.

Indeed, at high neutron energies ($E_n \geqslant 100~keV$), the second term in
the curly brackets of the numerator of the expression~(\ref{eq61}) is
significantly larger than the first, and so the energy spectrum of the
moderating neutrons in this part will coincide with the neutron fission
spectrum. This is confirmed by the results of the calculations presented in
Fig.~\ref{fig03} and coincides with the theoretical diagram shown in
Fig.~\ref{fig04}.

With the decrease of neutron energy ($ 10~eV \leqslant E_n \leqslant 100~keV$)
both terms in the curly brackets of the numerator in the
expression~(\ref{eq61}) become comparable, and therefore this part of the
energy spectrum of the moderating neutron may be called a "transition region",
because it is formed by the contributions of the two terms of the numerator in
the expression~(\ref{eq61}) -- i.e. the sum of the fission spectrum and the
Fermi spectrum ($\sim 1/E_n$). Please note that this part of the moderating
neutron spectrum is erroneously marked as the Fermi spectrum in the theoretical
scheme shown in Fig.~\ref{fig04}.

With further decrease of neutron energy ($E_n \leqslant 10~eV$) the first term
in the curly brackets of the numerator of~(\ref{eq61}) will be significantly
greater than the second one, and effectively determines the energy spectrum of
the moderating neutrons in this energy range. However, due to the term
$\frac{3}{2}kT$ in the numerator of~(\ref{eq61}), a neutron energy range of 
$\frac{3}{2}kT \ll E_n \leqslant 10~eV$ may be singled out -- the energy
spectrum of the moderating neutrons will coincide with the Fermi spectrum
($\sim 1/E_n$). This is confirmed by the results of the calculations presented
in Fig.~\ref{fig03} ($0.03~eV \ll E_n \leqslant 10~eV$). Let us note that
Fig.~\ref{fig04} shows the Fermi spectrum ($\sim 1/E_n$) up to a certain energy
$E_{threshold}$, below which the moderating neutron spectrum is given by the
Maxwell spectrum. The form of latter is defined by the temperature of the
neutron gas, which in its turn is calculated from the empirical formula
connecting it with the temperature of moderating medium.

With a further decrease of neutron energy we obtain a "transition" from the
Fermi spectrum to the low-energy spectrum near $\sim \frac{3}{2}kT$, and the
low-energy part of the moderating neutron spectrum $E_n \ll \frac{3}{2}kT$.

According to expression~(\ref{eq61}), the low-energy part of the neutron
spectrum must be constant, since the integral in the numerator of the first
term in the curly brackets of~(\ref{eq61}) almost does not change with the
energy decrease. However, it turns out that the microscopic elastic
cross-section grows exponentially towards low energies in this range (for
example, according to the ENDF/B-VII.0 data and~\cite{Stacey2001}, for hydrogen
the exponent increases in 1000 times (see. Fig.~\ref{fig05}), and for uranium
it increases in 100 times). Such behavior of the elastic scattering cross
section leads to the appearance of second maximum in the low-energy part of the
spectrum. So it is clear that the nature of this maximum is associated with the
moderation process of the non-equilibrium system of neutrons (emitted by an
isotropic source) on the thermalized system of the moderating medium nuclei.
Thus, it cannot be explained by a thermally equilibrium part of the neutron
system only, i.e. by Maxwell distribution.

In contrast to the above considerations, the analogous solution given
in~\cite{Feinberg1978,BartolomeyBat1989,Shirokov1998} was obtained for the
traditional scattering law, and the low-energy part of the spectrum has the
form of the Fermi spectrum ($\sim 1/E_n$). Consequently, it goes to infinity
with the neutron energy tending to zero, i.e. there is no low-energy maximum.
Therefore, in order to somehow fit the experimental data in the framework of
the traditional theory of neutron moderation, the Fermi spectrum ($\sim 1/E_n$)
is used to a certain boundary energy, below which the moderating neutrons
spectrum is given by the Maxwell spectrum, the form of which is defined by the
temperature of the neutron gas calculated by the empirical relation to the
temperature of the fissile medium (see~\nameref{sec01}).

\section{Neutron moderation in non-absorbing media with mass number $A>1$}
\label{sec05}

According to the neutron scattering law~(\ref{eq43}), which takes into account
the thermal motion of the moderating medium nuclei, the moderation law in
non-absorbing media with mass number $A>1$ is

\begin{equation}
P(E_1) dE_1 = 
\frac{dE_1}{\left[ E_{10}^{(L)} + \frac{1}{A} \cdot \frac{3}{2} kT \right] 
(1-\alpha)}
\label{eq63}
\end{equation}

For the moderation law~(\ref{eq63}) performing the calculations similar to
those in~\cite{Feinberg1978,BartolomeyBat1989,Shirokov1998}, we find the
following expression for the moderating neutrons flux density:

\begin{equation}
\Phi(E) = \frac{ \int \limits_{E} ^{\infty} Q(E') dE'}
{\left[ E + \frac{1}{A} \cdot \frac{3}{2} kT\right] \Sigma _s (E) \xi} + 
\frac{Q(E)}{\Sigma_s(E)} ,
\label{eq64}
\end{equation}

\noindent
where $Q(E)$ is the number of generated neutrons with an energy $E$ per unit
volume per unit time (see Section~\ref{sec03} above), $\xi$ is the mean
logarithmic energy decrement.

By analogy to the standard theory of neutrons moderation,
e.g.~\cite{Feinberg1978,BartolomeyBat1989,Shirokov1998}, for the mean
logarithmic energy decrement $\xi$ we introduce the following expression
(assuming $E_1 \neq 0$ and $E_0 \neq 0$):

\begin{equation}
\xi = \left \langle \ln{\frac{E_0}{E_1}} \right \rangle = 
\frac{\int \limits _{\alpha \left( E_0 + \frac{1}{A} \frac{3}{2}kT \right)} 
^{\left( E_0 + \frac{1}{A} \frac{3}{2}kT \right)} 
\ln{\frac{E_0}{E_1}} P(E_1) dE_1}
{\int \limits _{\alpha \left( E_0 + \frac{1}{A} \frac{3}{2}kT \right)} 
^{\left( E_0 + \frac{1}{A} \frac{3}{2}kT \right)} 
P(E_1) dE_1} =
\int \limits _{\alpha \left( E_0 + \frac{1}{A} \frac{3}{2}kT \right)} 
^{\left( E_0 + \frac{1}{A} \frac{3}{2}kT \right)} 
\ln{\frac{E_0}{E_1}} \frac{dE_1}{\left( E_0 + \frac{1}{A} \frac{3}{2} kT \right) 
(1-\alpha)}
\label{eq65}
\end{equation}

Let us perform the variable substitution $E_1 \Leftrightarrow x$ using the
relation

\begin{equation}
E_0 + \frac{1}{A} \cdot \frac{3}{2} kT = \frac{E_1}{x} ~~(x \neq 0),
\label{eq66}
\end{equation}

\noindent
which implies the following relations

\begin{equation}
E_0 = \frac{E_1}{x} - \frac{1}{A} \frac{3}{2}kT , ~~~ 
E_1 = x \left( E_0 + \frac{1}{A} \cdot \frac{3}{2}kT \right) , \nonumber
\end{equation}

\begin{equation}
x = \frac{E_1}{E_0 + \frac{1}{A} \frac{3}{2}kT}, ~~~
dx = \frac{dE_1}{E_0 + \frac{1}{A} \frac{3}{2}kT} , \nonumber
\end{equation}

\noindent
from which for~(\ref{eq65}) we obtain:

\begin{align}
\xi & = \int \limits _{\alpha} ^{1}
\ln {\left( \frac{\frac{E_1}{x} - \frac{1}{A} \frac{3}{2}kT}{E_1} \right)} \frac{dx}{(1-\alpha)} =
\frac{1}{(1-\alpha)} \int \limits _{\alpha}^{1}
\ln{\left( \frac{1}{x} - \frac{1}{E_1} \frac{1}{A} \frac{3}{2}kT \right)} dx = \nonumber \\
& = \frac{1}{(1-\alpha)} \int \limits _{\alpha}^{1}
\ln{\left( \frac{1}{x} - \frac{1}{x \left( E_0 + \frac{1}{A} \frac{3}{2}kT \right)} 
\frac{1}{A} \frac{3}{2}kT \right)} dx = \nonumber \\
& = \frac{1}{(1-\alpha)} \int \limits _{\alpha}^{1}
\ln{\left(\frac{1}{x} \cdot \frac{E_0}{ E_0 + \frac{1}{A} \frac{3}{2}kT} \right)} dx = \nonumber \\
& = \frac{1}{(1-\alpha)} \left[ \int \limits _{\alpha}^{1}
\ln{\frac{1}{x}} dx + \ln{ \frac{E_0}{ E_0 + \frac{1}{A} \frac{3}{2}kT}} 
\int \limits _{\alpha}^{1} dx \right] = \nonumber \\
& = \frac{1}{(1-\alpha)} \int \limits _{\alpha}^{1}
\ln{\frac{1}{x}} dx + \frac{1}{(1-\alpha)} \ln{ \frac{E_0}{ E_0 + \frac{1}{A} \frac{3}{2}kT}} 
\left( 1 - \alpha \right) = \nonumber \\
& = \frac{1}{(1-\alpha)} \int \limits _{1} ^{\alpha}
\ln{x} dx + \ln{ \frac{E_0}{ E_0 + \frac{1}{A} \frac{3}{2}kT}} .
\label{eq67}
\end{align}

Integrating by parts, we find that the integral in the expression~(\ref{eq67}):

\begin{align}
\int \limits _{1} ^{\alpha} \ln{x} dx & = 
x \ln {x} \vert _{1} ^{\alpha} - 
x \vert _{1} ^{\alpha} = \alpha \ln {\alpha} + \left( 1 - \alpha \right) .
\label{eq68}
\end{align}

Substituting~(\ref{eq68}) into~(\ref{eq67}) we obtain the final expression for
$\xi$:
\begin{equation}
\xi = \frac{\alpha}{1-\alpha} 
\ln{\alpha} + 1 + \ln {\left( \frac{E_0}{E_0 + \frac{1}{A} \frac{3}{2}kT} \right)} .
\label{eq69}
\end{equation}

\begin{figure}
\begin{center}
\includegraphics[width=16cm]{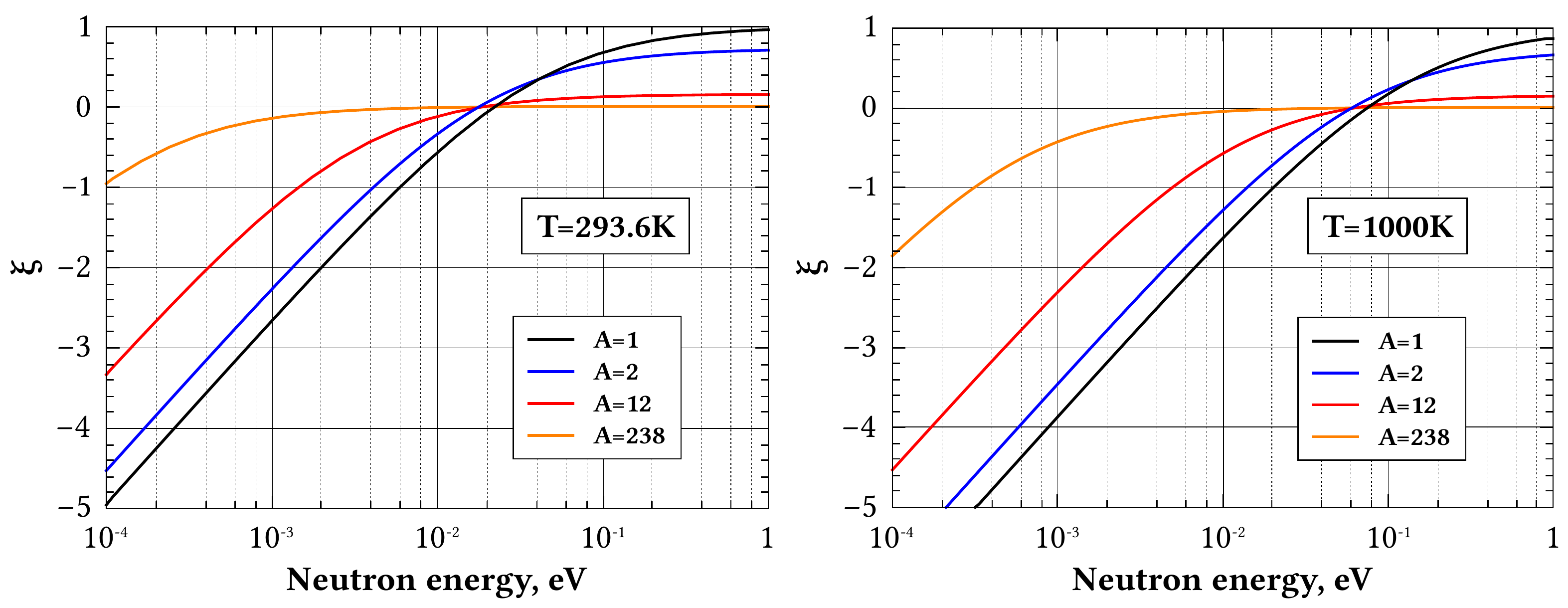}
\end{center}
\caption{Dependence of the logarithmic energy decrement on the neutron energy,
         calculated by~(\ref{eq69}).}
\label{fig06}
\end{figure}

Thus, in the framework of the new moderation theory we obtain the
expression~(\ref{eq69}) for $\xi$, according to which $\xi$ depends on the
initial energy of moderating neutrons $E_0$ and the moderating medium
temperature $T$. The dependence of the logarithmic energy decrement, calculated
by~(\ref{eq69}), is shown in Fig.~\ref{fig06}.

However, the expression~(\ref{eq69}) for 
$E_0 \gg \frac{1}{A} \cdot \frac{3}{2}kT$ turns into expression for $\xi$
obtained in the framework of the standard moderation theory,
e.g.~\cite{Feinberg1978,BartolomeyBat1989,Shirokov1998}:

\begin{equation}
\xi \approx 1 + \frac{\alpha}{1-\alpha} \ln{\alpha} = 
1 + \frac{(A-1)^2}{2A} \ln{\frac{A-1}{A+1}} .
\label{eq70}
\end{equation}

The approximate value of $\xi$ for heavy nuclei is

\begin{equation}
\xi \approx \frac{2}{A + \frac{2}{3}} .
\label{eq71}
\end{equation}

According to the obtained expression~(\ref{eq69}), for 
$E_0 \ll \frac{1}{A} \cdot \frac{3}{2}kT$ the mean logarithmic energy
decrement tends to zero with energy decrease, crosses zero, becoming negative
and tending further towards $-\infty$ (see Fig.~\ref{fig06}). The negative
values of $\xi$ correspond to the interactions of the neutrons with the medium
nuclei, in which neutrons gain additional energy. Such processes are not
considered by the standard moderation theory. It also means that one should
use $\vert \xi \vert$ instead of $\xi$ in the expression for the neutron
flux density~(\ref{eq64}), and take into account the zero crossing in the
calculations.


For the neutron flux density~(\ref{eq64}), as the neutron flux density
$\Phi(E)$ and the neutron density $n(E)$ are connected by~(\ref{eq60})
\cite{Feinberg1978,BartolomeyBat1989,Shirokov1998}, we find that the
probability density function of the moderating neutrons energy distribution is
given by the following expression:

\begin{equation}
\rho(E) = \frac{n(E)}{\int \limits_0 ^{\infty} n(E') dE'} = 
\frac{\frac{1}{\sqrt{2E/m_n}} 
\left \lbrace 
\frac{ \int \limits_{E} ^{\infty} Q(E') dE'}
{\left[ E + \frac{3}{2} kT\right] \Sigma _s (E)  \xi} + \frac{Q(E)}{\Sigma_s(E)}
\right \rbrace}
{ \int \limits _{0} ^{\infty} \left[
\frac{1}{\sqrt{2E'/m_n}}
\left \lbrace 
\frac{ \int \limits_{E'} ^{\infty} Q(E'') dE''}
{\left[ E' + \frac{3}{2} kT\right] \Sigma _s (E') \xi} + \frac{Q(E')}{\Sigma_s(E')}
\right \rbrace \right] dE'} .
\label{eq72}
\end{equation}

$Q(E)$ is given by the fission spectrum of a fissile nuclide (see
Section~\ref{sec03} above).

\section{Neutron moderation in non-absorbing moderating media containing several sorts of nuclides}
\label{sec06}

In this case the neutron scattering law is also given by~(\ref{eq63}). Carrying
out the calculations similar to those
in~\cite{Feinberg1978,BartolomeyBat1989,Shirokov1998}, we find the following
expression for the moderating neutrons flux density:

\begin{equation}
\Phi(E) = \frac{ \int \limits_{E} ^{\infty} Q(E') dE'}
{\left[ E + \frac{1}{A} \cdot \frac{3}{2} kT\right] \Sigma _s (E) \bar{\xi}} + 
\frac{Q(E)}{\Sigma_s(E)} ,
\label{eq73}
\end{equation}

\noindent
where $Q(E)$ is the number of generated neutrons with energy $E$ per unit
volume per unit time (see Section~\ref{sec03} above), $\bar{\xi}$ is the mean
logarithmic energy decrement averaged over all sorts of nuclei in the
moderating medium~\cite{Feinberg1978,BartolomeyBat1989,Shirokov1998}:

\begin{equation}
\bar{\xi} = \frac{\sum \limits_{i=1}^N \Sigma _{si} \xi_i}
{\sum \limits_{i=1}^N \Sigma _{si}} =
\frac{\sum \limits_{i=1}^N \Sigma _{si} \xi_i}
{\Sigma _s}
\label{eq74}
\end{equation}

Similarly to the above, the probability density function of the moderating neutrons energy distribution is:

\begin{equation}
\rho(E) = \frac{n(E)}{\int \limits_0 ^{\infty} n(E') dE'} = 
\frac{\frac{1}{\sqrt{2E/m_n}} 
\left \lbrace 
\frac{ \int \limits_{E} ^{\infty} Q(E') dE'}
{\left[ E + \frac{3}{2} kT\right] \Sigma _s (E) \bar{\xi}} + \frac{Q(E)}{\Sigma_s(E)}
\right \rbrace}
{ \int \limits _{0} ^{\infty} \left[
\frac{1}{\sqrt{2E'/m_n}}
\left \lbrace 
\frac{ \int \limits_{E'} ^{\infty} Q(E'') dE''}
{\left[ E' + \frac{3}{2} kT\right] \Sigma _s (E') \bar{\xi}} + \frac{Q(E')}{\Sigma_s(E')}
\right \rbrace \right] dE'} .
\label{eq75}
\end{equation}

\section{Neutron moderation in absorbing moderating media containing several sorts of nuclides}
\label{sec07}

In this case, the neutron scattering law is also given by~(\ref{eq63}).
Performing the calculations  similar to those
in~\cite{Feinberg1978,BartolomeyBat1989,Shirokov1998,RusovPNE2015}, we find the
following expression for the moderating neutrons flux density:

\begin{equation}
\Phi(E) = \left \lbrace 
\frac{ \int \limits_{E} ^{\infty} Q(E') dE'}
{\left[ E + \frac{1}{A} \frac{3}{2} kT\right] \Sigma _t (E) \bar{\xi}} + 
\frac{Q(E)}{\Sigma_t(E)} \right \rbrace
\cdot \exp \left \lbrace
- \int \limits_{E} ^{\infty}
\frac{ \Sigma_a (E') dE'}
{\left[ E' + \frac{1}{A} \frac{3}{2} kT\right] \Sigma _t (E') \bar{\xi}}
\right \rbrace ,
\label{eq76}
\end{equation}

\noindent
where $\Sigma _{si}$ is the macroscopic scattering cross-section for the
$i^{\text{th}}$ nuclide, $\Sigma_t = \sum \limits_i \Sigma_s^i + \Sigma_a^i$ is the
total macroscopic cross-section of the fissile material, 
$\Sigma_s = \sum \limits_i \Sigma_s^i$  is the total macroscopic scattering
cross-section of the fissile medium, $\Sigma_a$ is the macroscopic absorption
cross-section.

The Eq.~(\ref{eq76}) contains the expression for the probability function for
the neutrons to avoid the resonance absorption, which now also contains the
moderating medium temperature, in contrast to the standard moderation
theory~\cite{Feinberg1978,BartolomeyBat1989,Shirokov1998,RusovVANT2012}:

\begin{equation}
\varphi(E) = \exp \left \lbrace
- \int \limits_{E} ^{\infty}
\frac{ \Sigma_a (E') dE'}
{\left[ E' + \frac{1}{A} \cdot \frac{3}{2} kT\right] \Sigma _t (E') \bar{\xi}}
\right \rbrace ,
\label{eq77}
\end{equation}

Similarly to the above, the probability density function of the moderating neutrons energy distribution is:

\begin{equation}
\rho(E) = \frac{n(E)}{\int \limits_0 ^{\infty} n(E') dE'} = 
\frac{\frac{1}{\sqrt{2E/m_n}} 
\left \lbrace 
\frac{ \int \limits_{E} ^{\infty} Q(E') dE'}
{\left[ E + \frac{3}{2} kT\right] \Sigma _t (E) \bar{\xi}} + \frac{Q(E)}{\Sigma_t(E)}
\right \rbrace \cdot\varphi(E)}
{ \int \limits _{0} ^{\infty} \left[
\frac{1}{\sqrt{2E'/m_n}}
\left \lbrace 
\frac{ \int \limits_{E'} ^{\infty} Q(E'') dE''}
{\left[ E' + \frac{3}{2} kT\right] \Sigma _t (E') \bar{\xi}} + \frac{Q(E')}{\Sigma_t(E')}
\right \rbrace  \cdot \varphi(E') \right] dE'} .
\label{eq78}
\end{equation}

Analyzing the energy spectrum represented by~(\ref{eq78}), together with its
comparison to the complete scheme of the moderating neutrons spectrum rarely
found in literature~\cite{Vladimirov1986}, shows that the moderating neutrons
spectrum may be described adequately by a single expression, taking into
account the moderator temperature.

Let us emphasize that in Section~\ref{sec04} we omitted the consideration of the
resonance neutron absorption function~(\ref{eq77}) during the analysis of the
expressions for the flux density and spectrum of the moderating neutrons,
because we considered a non-absorbing moderator. This function will affect the
ratio of the amplitudes of two maxima in the moderating neutrons flux density
and spectrum. It will also reveal a fine resonant structure of the moderating
neutron spectrum in the regions of resonance energies (similar to those shown
in Fig.~\ref{fig04} from~\cite{Vladimirov1986}).

\section{Conclusion}

We obtained the analytical expression for the neutron scattering law for an
isotropic source of neutrons, which includes a temperature of the moderating
medium as a parameter in general case. The analytical expressions for the
neutron flux density and the spectrum of moderating neutrons, also depending on
the medium temperature were obtained as well.

As an example of the correct description of the moderating neutron spectrum by
the obtained analytical expressions, we present the calculated total energy
spectra (from~0 to 5~MeV) of neutrons moderated by the hydrogen medium at
temperatures of 1000~K. The fission energy spectrum of neutrons was used for an
isotropic neutron source. The calculated spectra are in good agreement with the
available experimental data and theoretical concepts of the neutron moderation
theory.

The obtained expressions for the moderating neutrons spectra create space for
the new interpretations of the physical nature of the processes that determine
the form of the neutron spectrum in the thermal region. We found the impact of
the elastic scattering cross-sections behavior on the formation of the
low-energy maximum in the moderating neutron spectrum It is clear that the
nature of this maximum is associated with the process of the non-equilibrium
neutron system (generated by an isotropic neutron source) moderation by a
thermalized system of moderator nuclei. Therefore it cannot be explained by the
thermalized part of the neutron system only, and thus by the Maxwell
distribution.

In conclusion it may be noted that the substantially different behavior of the
elastic scattering cross-sections for different moderating media (see e.g. 
ENDF/B-VII.0 or~\cite{Stacey2001}) opens the possibility for the experimental
studies of these cross-sections impact on the formation of the low-energy
maximum in the moderating neutron spectrum, as well as for the experimental
verification of the described analytical expressions.

\bibliographystyle{unsrtnat}
\bibliography{Tarasov-Moderation}

\begin{thebibliography}{28}
\providecommand{\natexlab}[1]{#1}
\providecommand{\url}[1]{\texttt{#1}}
\expandafter\ifx\csname urlstyle\endcsname\relax
  \providecommand{\doi}[1]{doi: #1}\else
  \providecommand{\doi}{doi: \begingroup \urlstyle{rm}\Url}\fi

\bibitem[Weinberg and Wigner(1958)]{Weinberg1961}
A.~M. Weinberg and E.~P. Wigner.
\newblock \emph{The Physical Theory of Neutron Chain Reactors}.
\newblock The University of Chicago Press, 1958.

\bibitem[Akhiezer and Pomeranchuk(2002)]{Akhiezer2002}
A.I. Akhiezer and I.Ya. Pomeranchuk.
\newblock \emph{Introduction into the theory of neutron multiplication systems
  (reactors)}.
\newblock IzdAT, Moscow, 2002.
\newblock in Russian.

\bibitem[Galanin(1960)]{Galanin1971}
A.D. Galanin.
\newblock \emph{Thermal reactor theory}.
\newblock Pergamon Press, New York, 1960.

\bibitem[Feinberg et~al.(1978)Feinberg, Shikhov, and Troyanskii]{Feinberg1978}
S.~M. Feinberg, S.B. Shikhov, and V.B. Troyanskii.
\newblock \emph{The theory of nuclear reactors, Volume 1, Elementary theory of
  reactors}.
\newblock Atomizdat, Moscow, 1978.
\newblock in Russian.

\bibitem[Bartolomey et~al.(1989)Bartolomey, Bat', Baibakov, and
  Altukhov]{BartolomeyBat1989}
G.G. Bartolomey, G.A. Bat', V.D. Baibakov, and M.S. Altukhov.
\newblock \emph{Basic theory and methods of nuclear power installation
  calculations}.
\newblock Energoatomizdat, Moscow, 1989.
\newblock in Russian.

\bibitem[Shirokov(1998)]{Shirokov1998}
S.V. Shirokov.
\newblock \emph{The nuclear reactor physics}.
\newblock Naukova Dumka, Kiev, 1998.
\newblock in Russian.

\bibitem[Stacey(2001)]{Stacey2001}
W.M. Stacey.
\newblock \emph{Nuclear Reactor Physics}.
\newblock Wiley-VCH, 2001.

\bibitem[Vladimirov(1986)]{Vladimirov1986}
V.I. Vladimirov.
\newblock \emph{Practical problems of nuclear reactor operation}.
\newblock Energoatomizdat, Moscow, 1986.
\newblock in Russian.

\bibitem[Verkhivker and Kravchenko(2008)]{Verkhivker2008}
G.P. Verkhivker and V.P. Kravchenko.
\newblock \emph{Bases for calculation and design of nuclear power reactors}.
\newblock TEC Publishing, Odessa, 2008.
\newblock in Russian.

\bibitem[Rusov et~al.(2011{\natexlab{a}})Rusov, Linnik, Tarasov, Zelentsova,
  Sharph, Vaschenko, Kosenko, Beglaryan, Chernezhenko, Molchinikolov, Saulenko,
  and Byegunova]{RusovEnergies2011}
Vitaliy~D. Rusov, Elena~P. Linnik, Victor~A. Tarasov, Tatiana~N. Zelentsova,
  Igor~V. Sharph, Vladimir~N. Vaschenko, Sergey~I. Kosenko, Margarita~E.
  Beglaryan, Sergey~A. Chernezhenko, Pavel~A. Molchinikolov, Sergey~I.
  Saulenko, and Olga~A. Byegunova.
\newblock Traveling wave reactor and condition of existence of nuclear burning
  soliton-like wave in neutron-multiplying media.
\newblock \emph{Energies}, 4\penalty0 (9):\penalty0 1337, 2011{\natexlab{a}}.
\newblock ISSN 1996-1073.
\newblock \doi{10.3390/en4091337}.
\newblock URL \url{http://www.mdpi.com/1996-1073/4/9/1337}.

\bibitem[Rusov et~al.(2011{\natexlab{b}})Rusov, Tarasov, and
  Chernezhenko]{RusovVANT2011}
V.D. Rusov, V.A. Tarasov, and S.A. Chernezhenko.
\newblock The modes with the sharpening in the uranium-plutonium fission
  environment of the technical nuclear reactors and georeactor.
\newblock \emph{Problems of Atomic Science and Technology}, 97\penalty0
  (2):\penalty0 123--131, 2011{\natexlab{b}}.
\newblock in Russian.

\bibitem[Rusov et~al.(2013{\natexlab{a}})Rusov, Tarasov, Vaschenko, Linnik,
  Zelentsova, Beglaryan, Chernegenko, Kosenko, Molchinikolov, Smolyar, and
  Grechan]{RusovWJNST2013}
V.~D. Rusov, V.~A. Tarasov, V.~M. Vaschenko, E.~P. Linnik, T.~N. Zelentsova,
  M.~E. Beglaryan, S.~A. Chernegenko, S.~I. Kosenko, P.~A. Molchinikolov, V.~P.
  Smolyar, and E.~V. Grechan.
\newblock Fukushima plutonium effect and blow-up regimes in neutron-multiplying
  media.
\newblock \emph{World Journal of Nuclear Science and Technology}, 3\penalty0
  (2A):\penalty0 9--18, 2013{\natexlab{a}}.
\newblock \doi{10.4236/wjnst.2013.32A002}.
\newblock arXiv:1209.0648v1.

\bibitem[Rusov et~al.(2013{\natexlab{b}})Rusov, Litvinov, Linnik, Vaschenko,
  Zelentsova, Beglaryan, Tarasov, Chernegenko, Smolyar, Molchinikolov,
  Merkotan, and Kavatskyy]{RusovJMP2013}
V.D. Rusov, D.A. Litvinov, E.P. Linnik, V.M. Vaschenko, T.N. Zelentsova, M.E.
  Beglaryan, V.A. Tarasov, S.A. Chernegenko, V.P. Smolyar, P.A. Molchinikolov,
  K.K. Merkotan, and P.~Kavatskyy.
\newblock Kamland-experiment and soliton-like nuclear georeactor. part 1.
  comparison of theory with experiment.
\newblock \emph{Journal of Modern Physics}, 4\penalty0 (4):\penalty0 528--550,
  2013{\natexlab{b}}.
\newblock \doi{10.4236/jmp.2013.44075}.

\bibitem[Rusov et~al.(2010)Rusov, Tarasov, Chernegenko, and
  Borikov]{RusovKhariton2010}
V.D. Rusov, V.A. Tarasov, S.A. Chernegenko, and T.L. Borikov.
\newblock Blow-up modes in uranium-plutonium fissile medium in technical
  nuclear reactors and georeactor.
\newblock In \emph{Proc. Int. Conf. "Problems of physics of high energy
  densities. XII Khariton Topical Scientific Readings"}, pages 94--102, Sarov,
  2010. Publishing "RFNC-VNIIEF".
\newblock in Russian.

\bibitem[Prigogine(1968)]{Prigogine2001}
I.~Prigogine.
\newblock \emph{Introduction to Thermodynamics of Irreversible Processes}.
\newblock John Wiley \& Sons, New York, 1968.

\bibitem[Bakhareva(1976)]{Bakhareva1976}
I.F. Bakhareva.
\newblock \emph{Nonlinear nonequilibrium thermodynamics}.
\newblock Saratov University Publishing, 1976.
\newblock in Russian.

\bibitem[Kvasnikov(2003)]{Kvasnikov2003}
I.A. Kvasnikov.
\newblock \emph{Thermodynamics and statistical physics. Vol.3 "Theory of
  nonequilibrium systems"}.
\newblock Editorial URSS, 2003.
\newblock in Russian.

\bibitem[Rusov et~al.(2015{\natexlab{a}})Rusov, Tarasov, Sharph, Vashchenko,
  Linnik, and1 M.~E.~Beglaryan, Chernegenko, Kosenko, and
  Smolyar1]{RusovSTNI2015}
V.~D. Rusov, V.~A. Tarasov, I.~V. Sharph, V.~N. Vashchenko, E.~P. Linnik,
  T.~N.~Zelentsova and1 M.~E.~Beglaryan, S.~A. Chernegenko, S.~I. Kosenko, and
  V.~P. Smolyar1.
\newblock On some fundamental peculiarities of the traveling wave reactor.
\newblock \emph{Science and Technology of Nuclear Installations},
  2015:\penalty0 703069, 2015{\natexlab{a}}.

\bibitem[Rusov et~al.(2015{\natexlab{b}})Rusov, Tarasov, Eingorn, Chernezhenko,
  Kakaev, Vashchenko, and Beglaryan]{RusovPNE2015}
V.D. Rusov, V.A. Tarasov, M.V. Eingorn, S.A. Chernezhenko, A.A. Kakaev, V.M.
  Vashchenko, and M.E. Beglaryan.
\newblock Ultraslow wave nuclear burning of uranium-plutonium fissile medium on
  epithermal neutrons.
\newblock \emph{Progress in Nuclear Energy}, 83:\penalty0 105--122,
  2015{\natexlab{b}}.

\bibitem[Kolesov(2006)]{Kolesov2007}
V.~F. Kolesov.
\newblock \emph{Aperiodic pulse reactors. Vol.1}.
\newblock RFNC-VNIIEF Publishing, Sarov, 2006.

\bibitem[Lukin(2006)]{Lukin2006}
Lukin.
\newblock \emph{Physics of the Pulse Nuclear Reactors}.
\newblock RFNC-VNIITF Publishing, Snezhinsk, 2006.
\newblock in Russian.

\bibitem[Arapov(2010)]{Arapov2010}
Arapov.
\newblock The results of the physical launch of the br-1m reactor.
\newblock In \emph{Problems of the physics of high energy density. XII
  Kharitonov thematic scientific readings}, pages 22--27, Sarov, 2010.
  RFNC-VNIIEF Publishing.
\newblock in Russian.

\bibitem[Rusov et~al.(2007)Rusov, Pavlovich, Vaschenko, Tarasov, Zelentsova,
  Bolshakov, Litvinov, Kosenko, and Byegunova]{RusovJGR2007}
V.~D. Rusov, V.~N. Pavlovich, V.~N. Vaschenko, V.~A. Tarasov, T.~N. Zelentsova,
  V.~N. Bolshakov, D.~A. Litvinov, S.~I. Kosenko, and O.~A. Byegunova.
\newblock Geoantineutrino spectrum and slow nuclear burning on the boundary of
  the liquid and solid phases of the earth's core.
\newblock \emph{Journal of Geophysical Research: Solid Earth}, 112\penalty0
  (B9):\penalty0 n/a--n/a, 2007.
\newblock ISSN 2156-2202.
\newblock \doi{10.1029/2005JB004212}.
\newblock URL \url{http://dx.doi.org/10.1029/2005JB004212}.
\newblock B09203.

\bibitem[Levich et~al.(1973)Levich, Myamlin, and Vdovin]{Levich1971}
Benjamin~G. Levich, V.A. Myamlin, and Yu.A. Vdovin.
\newblock \emph{Theoretical Physics: An Advanced Text Volume 3: Quantum
  Mechanics}.
\newblock John Wiley \& Sons, New York, 1973.

\bibitem[Levich(1971)]{Levich1969}
V.G Levich.
\newblock \emph{Theoretical Physics: An Advanced Text, Vol. 2: Statistical
  Physics, Electromagnetic Processes in Matter}.
\newblock Elsevier Science Publishing Co Inc., U.S., 1971.

\bibitem[Fedorov(1961)]{EngPhysRef1961}
N.D. Fedorov.
\newblock \emph{A brief reference book for engineer-physicists}.
\newblock State publishing of literature in the field of nuclear science and
  technology, Moscow, 1961.
\newblock in Russian.

\bibitem[Shirokov and Yudin(1983)]{Shirokov1972}
Yu.~M. Shirokov and N.P. Yudin.
\newblock \emph{Nuclear Physics}.
\newblock Imported Pubn, 1983.

\bibitem[adn V.A.~Tarasov et~al.(2012)adn V.A.~Tarasov, Kosenko, and
  Chernezhenko]{RusovVANT2012}
V.D.~Rusov adn V.A.~Tarasov, S.I. Kosenko, and S.A. Chernezhenko.
\newblock The function of resonance absorption probability for the neutron and
  multiplicate integral.
\newblock \emph{Problems of Atomic Science and Technology}, 99\penalty0
  (2):\penalty0 68--72, 2012.
\newblock in Russian.

\end{thebibliography}

\end{document}